\documentclass[10pt, twocolumn,twoside]{IEEEtran} 
\IEEEoverridecommandlockouts                              

\usepackage{color}
  \usepackage{algorithmicx}
    \usepackage[ruled]{algorithm}
    \usepackage{algpseudocode}
\usepackage{graphics} 
\usepackage{epsfig} 
\usepackage{subfigure}
\usepackage{ulem}
\usepackage{times} 
\usepackage{amsmath} 
\usepackage{amssymb}  
\usepackage{tikz}

\newtheorem{theorem}{Theorem}[section]

\newtheorem{definition}{Definition}[section]

\newtheorem{example}[definition]{Example}

\newtheorem{remarkth}[definition]{Remark}
\newenvironment{remark}{\begin{remarkth}\upshape}{\end{remarkth}}

\usepackage{amssymb}

\newcommand{\proa}{A^*G \mbox{$\;$}_{\tau^*} \kern-3pt\times_\alpha
G \mbox{$\;$}_\beta \kern-3pt\times_{\tau^*} A^*G}

\newcommand{\Ad}{\mbox{Ad}}

\newcommand{\ad}{\mbox{ad}}

\tikzstyle{vertex}=[circle,fill=black!20,minimum size=15pt,inner sep=0pt]
\tikzstyle{selected vertex} = [vertex, fill=red!24]
\tikzstyle{edge} = [draw,thick,-]
\tikzstyle{dedge} = [draw,thick,<->]
\tikzstyle{shadowdedge} = [draw, dotted,->]
\tikzstyle{weight} = [font=\small]
\tikzstyle{selected edge} = [draw,line width=3pt,-,red!50]
\tikzstyle{ignored edge} = [draw,line width=3pt,-,black!20]

\usepackage{nomencl}
\usepackage{ifthen}
\renewcommand{\nomgroup}[1]{%
\ifthenelse{\equal{#1}{C}}{\item[\textbf{Constants}]}{%
\ifthenelse{\equal{#1}{V}}{\item[\textbf{Variables}]}{%
\ifthenelse{\equal{#1}{S}}{\item[\textbf{sets}]}{}}}
}
\makenomenclature

\title{\LARGE \bf
Motion Feasibility Conditions for\\ Multi-Agent Control Systems on Lie Groups}

\author{Leonardo J. Colombo and Dimos V. Dimarogonas, \textit{Senior Member, IEEE}

\thanks{L. J. Colombo (leo.colombo@icmat.es) is with Instituto de Ciencias Matem\'aticas (CSIC-UAM-UCM-UC3M), Calle Nicol\'as Cabrera 13-15, Campus Cantoblanco, 28049, Madrid, Spain. D.V. Dimarogonas (dimos@kth.se) is with Department of Automatic Control, EECS School, KTH, Royal Institute of Technology, SE-100 44, Stockholm, Sweden.} 
\thanks{The work of Leonardo J. Colombo was also partially supported by ACCESS Linnaeus Center, KTH, Royal Institute of Technology, Sweden;  Ministerio de Econom\'ia, Industria y
Competitividad (MINEICO, Spain) under grant MTM2016-76702-P, Juan de la Cierva Incorporaci\'on fellowship and I-Link Project \textit{linkA20079}. The project that gave rise to these results received the support of a fellowship from ''la Caixa'' Foundation (ID 100010434). The fellowship code is LCF/BQ/PI19/11690016. 

The work of Dimos V. Dimarogonas is supported by the Swedish Research Council (VR), Knut och Alice Wallenberg foundation (KAW), the H2020 Project Co4Robots and the H2020 ERC Starting Grant BUCOPHSYS.}}%
\begin{document}

\maketitle
\thispagestyle{empty}
\pagestyle{empty}

\begin{abstract}
We study the problem of motion feasibility for multi-agent control systems on Lie groups with collision avoidance constraints. We first consider the problem for kinematic left invariant control systems and next, for dynamical control systems given by a left-trivialized Lagrangian function. Solutions of the kinematic problem give rise to linear combinations of the control inputs in a linear subspace annihilating the collision avoidance constraints. In the dynamical problem, motion feasibility conditions are obtained by using techniques from variational calculus on manifolds, given by a set of equations in a vector space, and Lagrange multipliers annihilating the constraint force that prevents deviation of solutions from a constraint submanifold. 

  \end{abstract}
\begin{IEEEkeywords} Mechanical systems on Lie groups, Collision avoidance, Multi-agent systems, Variational principles, Left-invariant control systems. \end{IEEEkeywords}
\section{Introduction}

Decentralized control strategies for multiple vehicles have gained increased attention in the last decades in the control community \cite{jorgebook}, \cite{MMbook}, \cite{oh}. In particular, when the configuration space of the agents is on a Lie group the main applications involved the coordination and synchronization of spacecraft motions modeled by kinematic systems \cite{SBS}, \cite{SaLS}. Recently, researchers have shown an inceinterest in employing decentralized motion planning algorithms for multi-agent systems based on second-order dynamical models \cite{dimos2}, \cite{magnus}. The main motivation lies in that acceleration controls are more implementable in vehicle systems than velocity controls.

In this work we consider a set of agents evolving on a Lie group subject to collision avoidance constraints. We determine whether there are non-trivial trajectories in the collision avoidance problem of all agents that maintain the constraints. The results are applied to build a collision avoidance motion planning controller for the coordinated motion of the agents. We assume  that  the  constraints  for  the  distributed  edge set  should  be  non-conflicting,  and  the  overall  constraint  for  all  the edges  should  be  realizable  in  the  full  Lie group. 
The proposed mathematical framework for multi-agent systems on Lie groups was recently used in \cite{CoDi} for optimal control problems. We also build on the works \cite{tabuada}, \cite{sun} by studying the problem of motion feasibility when agents evolves on Lie group manifolds.

The motion feasibility problem is studied in two different scenarios:  when agents are described by kinematic left-invariant fully actuated control systems and when the agents are described by dynamical fully actuated control systems. While the kinematic approach has been studied more in the literature for the motion feasibility problem (single integrator dynamics), the main motivation for the second approach lies in the fact that acceleration control (double integrator dynamics) is more suitable under the real world requirements of sensors for multiple vehicles, than velocity controls. It also provides a first step towards the construction of distance-based numerical estimators via Lie groups variational integrators \cite{CoGdM}.   The solution in the second approach is given by using techniques of calculus of variations on manifolds and the Lagrange multiplier theorem, while for the first one, we use techniques of differential calculus on manifolds.

The main contribution of this work consists on providing a set of necessary conditions for non-trivial collision-free motions in multi-agent control systems where agents evolves on a Lie group manifold. The main results of this work are given in Theorem \ref{theorem} and Theorem \ref{theorem1var}. Theorem \ref{theorem} describes differential-algebraic conditions for the feasible motion, when the agents are given by kinematic left-invariant control systems,  by finding the set of admissible velocities leaving the constraints invariant at a given point on a Lie group $G$ and describing it as a linear system of algebraic equations with the control inputs as unknown variables. Theorem \ref{theorem1var}, provides first-order necessary conditions for feasible motion when the dynamics of each agent is described by a Lagrangian function on $G\times\mathfrak{g}$ through the constrained Euler-Lagrange equations,  with $\mathfrak{g}$ being the Lie algebra associated with $G$. Such a condition is given by a set of first order differential equations on $\mathfrak{g}$. 

The paper is structured as follows. Section II provides the nomenclature. Section III  introduces Lie groups actions, constrained Euler-Lagrange equations and trivializations of the tangent bundle of a Lie group. Section IV describes left-invariant kinematic multi-agent control systems, dynamical multi-agent control systems and the formulation for the motion feasibility problem.  In Section V we consider a differential-algebraic approach for the motion feasibility problem of kinematic left-invariant multi-agent systems. In Section VI we derive first-order necessary conditions for feasible motion through constrained Euler-Lagrange equations arising from a variational point of view. Section VII studies the applicability of the conditions found in Sections V and VI for the collision avoidance problem of three rigid bodies on $SE(3)$ modelling fully-actuated underwater vehicles. 

\section{Nomenclature}

We begin by establishing the nomenclature used throughout this
paper. The basic notation and methodology is fairly standard
within the differential geometry literature and we have attempted
to use traditional symbols and definitions wherever
feasible. Table \ref{tab:table1} provides the symbols
will be used frequently along the paper.

\begin{table}[h!]
  \begin{center}
    \caption{Nomenclature}
    \label{tab:table1}
    \begin{tabular}{l|c} 
      \textbf{Symbol} & \textbf{Description} \\
      \hline
      $Q$ & Differentiable manifold\\
      $TQ$  & Tangent bundle of $Q$ \\
      $T^{*}Q$ &  Cotangent bundle of $Q$\\
      $G$ & Lie group\\
       $\mathfrak{g}$ & Lie algebra of $G$\\ 
       $\mathfrak{g}^{*}$ & Dual of the Lie algebra $\mathfrak{g}$\\
       $n$ & Dimension of $G$\\
       $r$ & Number of agents\\
       $p$ & Number of edges in the communication graph\\
       $T$ & Transpose of a Matrix\\
       $\overline{m}$ & Quantity of collision avoidance constraints\\
       $\lambda_k$ & Lagrange multiplier\\
       $\phi_{ij}^{k}:G\times G\to\mathbb{R}$ & Collision avoidance constraints\\
       $\lambda_{TG}$ & Left-trivialization of $TG$\\
       $\lambda_{T^{*}G}$& Left-trivialization of $T^{*}G$\\
       $g,h$ & Elements on $G$\\
       $\xi$ & Element on $\mathfrak{g}$\\ 
       $\mu$ & Element of $\mathfrak{g}^{*}$\\
       $\overline{e}$&  Identity element of $G$\\
       $L_g:G\to G$ & Left-translation $L_gh=gh$\\
       $T_hL_g$ &tangent map of $L_g$ at $h \in G$\\
       $T^{*}_hL_g$ & the cotangent map of $L_g$ at $h \in G$\\
       $\Ad:G\times\mathfrak{g}\to\mathfrak{g}$ & Adjoint action\\ 
       $\Ad^{*}:G\times\mathfrak{g}^{*}\to\mathfrak{g}^{*}$ &co-Adjoint action\\
     $\ad:\mathfrak{g}\times\mathfrak{g}\to\mathfrak{g}$ & Adjoint operator\\
$\ad^{*}:\mathfrak{g}\times\mathfrak{g}^{*}\to\mathfrak{g}^{*}$ & co-Adjoint operator
    \end{tabular}
  \end{center}
\end{table}
\section{Preliminaries}

\subsection{Differential calculus on manifolds}

Let $Q$ be a differentiable manifold with $\dim(Q)=n$. Given a tangent vector $v_q\in T_qQ$, $q=(q^1,\ldots,q^n)\in Q$, and $f\in\mathcal{C}^{\infty}(Q)$, the set of real valued smooth functions on $Q$, $df\cdot v_q$  denotes how tangent vectors acts on functions on $\mathcal{C}^{\infty}(Q)$. $df$ denotes the differential of the function $f\in\mathcal{C}^{\infty}(Q)$ defined as $$df(q)\cdot v_q=\sum_{i=1}^n\frac{\partial f}{\partial q^i}\cdot v_q^i.$$

Just as a vector field is a ``field'' for tangent vectors, a differential  $1$-form is a ``field'' of cotangent vectors, one for every base point.  A cotangent vector based at $q\in Q$, is a linear map from $T_{q}Q$ to $\mathbb{R}$, and the set of all maps is the cotangent space $T^{*}_{q}Q$, which is the dual to the tangent space $T_{q}Q$. A $1$-form on $Q$ is a map $\Theta:Q\to T^{*}Q$ such that $\Theta(q)\in T^{*}Q$ for every $q\in Q$. Differential one forms, can be added together and multiplied by scalar fields $c:Q\to\mathbb{R}$ as $(\Theta+\overline{\Theta})(q)=\Theta(q)+\overline{\Theta}(q)$, and $(c\Theta)(q)=c(q)\Theta(q)$.

Given a differentiable function $f:Q\to Q_1$ with $Q_1$ a smooth manifold, the pushforward of $f$ at $q\in Q$ is the linear map $T_{q}f:T_qQ\to T_{f(q)}Q_1$ satisfying $T_qf(v_q)\cdot\phi=v_q\cdot d(\phi\circ f)$ for all $\phi\in\mathcal{C}^{\infty}(Q_1)$ and $v_q\in T_qQ$. The pullback of $f$ at $q\in Q$ is the dual map $T^{*}_{q}f:T^{*}_{f(q)}Q_1\to T^{*}_qQ$ satisfying \begin{equation}\label{pairingdual}\langle T^{*}_{q}f(p_q),v_q\rangle_{*}=\langle p_q,T_{q}f(v_q)\rangle_{*}\end{equation} for all $v_q\in T_{q}Q$ and $p_q\in T^{*}_{f(q)}Q_1$, where $\langle \cdot,\cdot\rangle_{*}$ denotes how tangent covectors acts on tangent vectors.

\begin{definition}
Let $Q$ and $N$ be differentiable manifolds and $f:Q\to N$ be a differentiable map between them. The map $f$ is a submersion at a point $q\in Q$ if its differential $df(q):T_q Q \to T_{f(q)}N$ is a surjective map.\end{definition}

\begin{definition}Let $U\subset Q$ be an open set, $f:U\to N$ be smooth. If $f$ is a submersion at all points in $U$ then for all $y\in N$, $f^{-1}(y)\subset U$ is a submanifold of $Q$. The value $y\in N$ is said to be a regular value of $f$. \end{definition}

\begin{theorem}\label{LMth}[\cite{MaRa1999}, Section $8.3$, pp. 219]
Let $Q$ be a differentiable manifold and $\mathbf{0}\in\mathbb{R}^{m}$ a regular value of $\Phi:Q\to\mathbb{R}^{m}$. Given a function $\mathcal{S}:Q\to\mathbb{R}$, by defining the function $\overline{\mathcal{S}}:Q\times\mathbb{R}^{m}\to\mathbb{R}$ as $$\overline{\mathcal{S}}(q,\lambda)=\mathcal{S}(q)-\langle\langle\lambda,\Phi(q)\rangle\rangle,$$ for some inner product $\langle\langle\cdot,\cdot\rangle\rangle$ on $\mathbb{R}^{m}$, the Lagrange multiplier theorem states that $q\in \Phi^{-1}(\mathbf{0})$ is an extrema of $\mathcal{S}\mid_{\Phi^{-1}(\mathbf{0})}$ if and only if $(q,\lambda)\in Q\times\mathbb{R}^{m}$ is an extrema of $\overline{\mathcal{S}}$. 
\end{theorem}

\subsection{Lie group actions}\label{section2}

\begin{definition} Let $G$ be a Lie group with the identity element $\bar{e}\in G$. A \textit{left-action} of $G$ on a manifold $Q$ is a smooth mapping $\Phi:G\times Q\to Q$ such that $\Phi(\overline{e},q)=q$ for all $q\in Q$, $\Phi(g,\Phi(h,q))=\Phi(gh,q)$ for all $g,h\in G, q\in Q$ and for every $g\in G$, the map $\Phi_g:Q\to Q$ defined by $\Phi_{g}(q):=\Phi(g,q)$ is a diffeomorphism. \end{definition}

Let $G$ be a finite dimensional Lie group. The tangent bundle at a point $g\in G$ is denoted as $T_{g}G$ and the cotangent bundle at a point $h\in G$ is denoted as $T_{h}^{*}Q$. $\mathfrak{g}$ will denote the Lie algebra associated to $G$ defined as $\mathfrak{g}:=T_{\overline{e}}G$, the tangent space at the identity $\overline{e}\in G$. 
Given that the Lie algebra $\mathfrak{g}$ is a vector space, one may consider its dual space. Such dual of the Lie algebra is denoted by $\mathfrak{g}^{*}$.

 Let $L_{g}:G\to G$ be the left translation of the element $g\in G$ given by $L_{g}(h)=gh$ for $h\in G$. $L_g$ is a diffeomorphism on $G$ and a left-action from $G$ to $G$. Their tangent map  (i.e, the linearization or tangent lift of left translations) is denoted by $T_{h}L_{g}:T_{h}G\to T_{gh}G$. Similarly, the cotangent map (cotangent lift of left translations) is denoted by $T_{h}^{*}L_{g}:T^{*}_{h}G\to T^{*}_{gh}G$. It is known that the tangent and cotangent lift are actions (see \cite{Holmbook}, Chapter $6$).

Consider the vector bundles isomorphisms $\lambda_{TG}:G\times\mathfrak{g}\to TG$ and $\lambda_{T^*G}:G\times \mathfrak{g}^{*}\to T^{*}G$ defined as \begin{equation}\label{left-trivialization}\lambda_{TG}(g,\xi)=(g,T_{\overline{e}}L_{g}(\xi)),\,\,\lambda_{T^{*}G}(g,\mu)=(g,T_{g}^{*}L_{g^{-1}}(\mu)).\end{equation} $\lambda_{TG}$ and $\lambda_{T^{*}G}$ are called left-trivializations of $TG$ and $T^{*}G$ respectively. Therefore, the left-trivialization $\lambda_{TG}$ permits to identify  tangent bundle $TG$ with $G\times\mathfrak{g}$, and by $\lambda_{T^{*}G}$, the cotangent bundle $T^*{}G$ can be identified with $G\times\mathfrak{g}^{*}$. 

\begin{definition}[\cite{Holmbook}, Section $2.3$ pp.$72$]
The \textit{natural pairing} between vectors and co-vectors $\langle\cdot,\cdot\rangle:\mathfrak{g}^{*}\times\mathfrak{g}\to\mathbb{R}$ is defined by $\langle\alpha,\beta\rangle:=\alpha\cdot\beta$ for $\alpha\in\mathfrak{g}^{*}$ and $\beta\in\mathfrak{g}$ where $\alpha$ is understood as a row vector and $\beta$ a column vector. For matrix Lie algebras $\langle\alpha,\beta\rangle=\alpha^{T}\beta$. 
\end{definition}

Using the pairing between vectors and co-vectors, one can write the relation between the tangent and cotangent lifts as \begin{equation}\label{relation-cltl}\langle\alpha,T_{h}L_{g}(\beta)\rangle=\langle T^{*}_{h}L_{g}(\alpha),\beta\rangle\end{equation} for $g,h\in G$, $\alpha\in\mathfrak{g}^{*}$ and $\beta\in\mathfrak{g}$.

Let $\Phi_g:Q\to Q$ for any $g\in G$ a left action on $G$; a function $f:Q\to\mathbb{R}$ is said to be \textit{invariant} under the action $\Phi_g$, if $f\circ\Phi_g=f$, for any $g\in G$. The \textit{Adjoint action}, denoted $\hbox{Ad}_{g}:\mathfrak{g}\to\mathfrak{g}$ is defined by $\hbox{Ad}_{g}\chi:=g\chi g^{-1}$ where $\chi\in\mathfrak{g}$. Note that this action represents a change of basis on $\mathfrak{g}$

The \textit{co-adjoint operator} $\ad^{*}:\mathfrak{g}\times\mathfrak{g}^{*}\to\mathfrak{g}^{*}$, $(\xi,\mu)\mapsto\ad^{*}_{\xi}\mu$ is defined by $\langle\ad_{\xi}^{*}\mu,\eta\rangle=\langle\mu,\ad_{\xi}\eta\rangle$ for all $\eta\in\mathfrak{g}$ with $\ad:\mathfrak{g}\times\mathfrak{g}\to\mathfrak{g}$ the \textit{adjoint operator} given by $\ad_{\xi}\eta:=[\xi,\eta]$  where $[\cdot,\cdot]$ denotes the Lie bracket of vector fields on $\mathfrak{g}$.

The co-Adjoint action $\Ad^{*}_{g^{-1}}:\mathfrak{g}^{*}\to\mathfrak{g}^{*}$ is given by $\langle\Ad^{*}_{g^{-1}}\mu,\xi\rangle=\langle\mu,\Ad_{g^{-1}}\xi\rangle$ with $\mu\in\mathfrak{g}^{*}$, $\xi\in\mathfrak{g}$. Note that $\Ad$ and $\Ad^{*}$ are actions on Lie groups, while $\ad$ and $\ad^{*}$ are operators on the Lie algebra and its dual, respectively.
\begin{example}\label{coadjoinsSO(3)}
The co-Adjoint action of $SO(3)$ on $\mathfrak{so}(3)^{*}$, $\Ad^{*}:SO(3)\times\mathfrak{so}(3)^{*}\to\mathfrak{so}(3)^{*}$ is given by (see \cite{Holmbook}, Ch. $6$, pp. 224) $\Ad_{R^{-1}}^{*}\breve{\eta}=(R\eta)^{\breve{\empty}}$, and identifying $\mathfrak{so}(3)^{*}$ with $\mathbb{R}^{3}$, it is given by $\Ad^{*}_{R^{-1}}\eta=R\eta$ with $\eta\in\mathbb{R}^{3},\breve{\eta}\in\mathfrak{so}(3)^{*}$ and $R\in SO(3)$.\hfill$\diamond$. 
\end{example}

Let $\mathcal{L}:TG\to\mathbb{R}$ be a Lagrangian function describing the dynamics of a mechanical system. After a left-trivialization of $TG$ we may consider the trivialized Lagrangian $\mathbf{L}:G\times\mathfrak{g}\to\mathbb{R}$ given by $\mathbf{L}(g,\xi)=\mathcal{L}(g,T_{g}L_{g^{-1}}(\dot{g}))=\mathcal{L}(g,g^{-1}\dot{g})$.

The left-trivialized Euler--Lagrange equations on $G\times\mathfrak{g}$ (see, e.g., \cite{Holmbook}, Ch. $7$), are given by the system of $n$ first order ode's \begin{equation}
\frac{d}{dt}\frac{\partial\mathbf{L}}{\partial\xi} +T^{*}_{\overline{e}}L_{g}\left(\frac{\partial\mathbf{L}}{\partial g}\right)=\ad^{*}_{\xi}\frac{\partial\mathbf{L}}{\partial\xi}\label{eq_ep_intro} \end{equation} together with the kinematic equation $\dot{g}=T_{\overline{e}}L_{g}\xi$, i.e., $\dot{g}=g\xi$.

The left-trivialized Euler-Lagrange equations together with the equation $\xi = T_{g}L_{g^{-1}}(\dot{g})$ are equivalent to the Euler--Lagrange equations for $\mathcal{L}$. Note that for a matrix Lie group, the previous equations means $\xi=g^{-1}\dot{g}$.

If $\mathbf{L}$ does not depend on $g\in G$ (for instance, as the Lagrangian for Euler's equations on the Lie group  $SO(3)$), equations \eqref{eq_ep_intro} reduce to the Euler-Poincar\'e equations \begin{align}
\frac{d}{dt}\frac{\partial\mathbf{L}}{\partial\xi} =\ad^{*}_{\xi}\frac{\partial\mathbf{L}}{\partial\xi}\label{eq_ep_intro2}\end{align} together with the kinematic equation $\dot{g}=T_{\overline{e}}L_{g}\xi$, i.e., $\dot{g}=g\xi$.

\section{Multi-agent control system on Lie groups}
In this section we introduce multi-agent control systems where the configuration space of each agent is a Lie group. It is described by an undirected static and connected graph. First, we introduce the motion feasibility problem of agents where each node of the graph is given by a dynamical control system (i.e., by the controlled trivialized Euler-Lagrange equations) governed by a Lagrangian function, and next, we consider that each node is given by a left-invariant control system.
\subsection{Left-invariant dynamical  multi-agent control systems}\label{lim}

Consider a set $\mathcal{N}$ consisting of $r$ free agents evolving each one on a Lie group $G$ with dimension $n$.  Along this work we assume that the configuration space of each agent has the same Lie group structure. Same configuration does not mean the same agent. For instance, each agent can have different masses and inertia values, and therefore agents are heterogeneous. 
We denote by $g_i\in G$ the configuration (positions) of an agent $i\in\mathcal{N}$ and $g_i(t)\in G$ describes the evolution of agent $i$ at time $t$. The element $g\in G^{r}$ denotes the stacked vector of positions where $G^{r}:=\underbrace{G\times\ldots\times G}_{r-times}$ represents the cartesian product of $r$ copies of $G$. We also consider  $\mathfrak{g}^{r}:=T_{\overline{e}}G^{r}$ the Lie algebra associated with the Lie group $G$ for the agent $i\in\mathcal{N}$ where $\overline{e}=(\overline{e}_1,\ldots,\overline{e}_r)$ is the neutral element of $G^{r}$ with $\overline{e}_j$ the neutral element of the $j^{th}$-Lie group which determines $G^{r}$.

The neighbor relationships are described by an undirected static and connected graph $\mathcal{G}=(\mathcal{N},\mathcal{E})$ where the set $\mathcal{E}\subset\mathcal{N}\times\mathcal{N}$ denotes the set of ordered edges for the graph. The set of neighbors for agent $i$ is defined by $\mathcal{N}_i=\{j\in\mathcal{N}: (i,j)\in\mathcal{E}\}$. 

The dynamics of each agent $i\in\mathcal{N}$ is determined by a Lagrangian function $\mathcal{L}_i:TG\to\mathbb{R}$ together with collision avoidance (holonomic) constraints. Each tangent space $TG$ can be left-trivialized and therefore, instead of working with $\mathcal{L}_i:TG\to\mathbb{R}$ we shall consider $\ell_i:G\times\mathfrak{g}\to\mathbb{R}$. Note also that the left-trivialization is not an extra assumption, we can always identify $TG$ with $G\times\mathfrak{g}$ by using the isomorphism \eqref{left-trivialization}.

Each agent $i\in\mathcal{N}$ is assumed to be a fully-actuated dynamical Lagrangian control system associated with the Lagrangian $\ell_i:G\times\mathfrak{g}\to\mathbb{R}$, that is,

 \begin{equation}\label{eq:left-invariant-dynamical-system-multi-agents}\frac{d}{dt}\frac{\partial\ell_i}{\partial\xi^{i}} -\ad^{*}_{\xi^{i}}\frac{\partial\ell_i}{\partial\xi^{i}}+T^{*}_{\overline{e}_i}L_{g_i}\left(\frac{\partial \ell_{i}}{\partial g_i}\right)=u_i,
\quad i\in\mathcal{N}\end{equation} where for each $i$, the $n$-tuple of control inputs $u_i=[u^{1}_i\ldots u^{n}_i]^{T}$ take values in $\mathbb{R}^{n}$ and where $g_i(\cdot)\in C^{1}([a_i,b_i],G)$ with $a_i,b_i\in\mathbb{R}^{+}$, and where $[a_i,b_i]$ is an arbitrary interval of $\mathbb{R}$.

We also assume that each agent $i\in\mathcal{N}$ occupies a disk of radius $\overline{r}$ on $G$. The quantity $\overline{r}$ is chosen to be small enough so that it is possible to pack $r$ disks of radius $\overline{r}$ on $G$. We say that agents $i$ and $j$ avoid mutual collision if $d(\pi_i(g),\pi_j(g))>\bar{r}$ where  $\pi_{i}:G^r\to G$ is the canonical projection from $G^r$ over its $i^{th}$-factor and $d$ is an appropriated distance function on the Lie group $G$.

Consider the set $\mathcal{C}$ given by the (holonomic or position-based) constraints indexed by the edges set $\mathcal{C}_{\mathcal{E}}=\{e_1,\ldots,e_p\}$ with $p=|\mathcal{E}|$, the cardinality of the set of edges. Each $e_{\alpha}\in\mathcal{C}_{\mathcal{E}}$ for $\alpha=1,\ldots,p$ is a set of constraints for the edge $e_{\alpha}=(i,j)\in\mathcal{E}$, that is, $e_{\alpha}=\{\phi_{ij}^1,\ldots,\phi_{ij}^{k_{\alpha}}\},$ being $k_{\alpha}$ the number of constraints on the edge $e_{\alpha}$. Let $\overline{m}$ be the total number of constraints in the set $\mathcal{N}$, that is, \begin{equation}\label{mbar}\displaystyle{\overline{m}=\sum_{\alpha=1}^{p}k_{\alpha}}.\end{equation} For each edge $e_{\alpha}$, $\phi_{ij}^{k}$ is a function on $G\times G$ defining an inter-agent collision avoidance constraint between agents $i$ and $j$ for all $k=1,\ldots,k_{\alpha}$ . The constraint is enforced if and only if $\phi_{ij}^{k}(g_i,g_j)=0$.

The constraints on edge $e_{\alpha}$, induce the constraints $\Phi_{ij}^{k}:G^r\to\mathbb{R}$ as $\Phi_{ij}^{k}(g)=\phi_{ij}^{k}(\pi_{i}(g),\pi_j(g))$.  If the map $\Phi:G^r\to\mathbb{R}^{m}$
is a submersion at any point of its domain, then $\mathcal{M}=\Phi^{-1}(0)$ is an $(rn-\overline{m})$-dimensional submanifold of $G^r$. Its $2(rn-\overline{m})$-dimensional tangent bundle is given by 
\begin{equation}\label{tm}T\mathcal{M}=\{(g,\dot{g})\in T_{g}G^r\mid\Phi(g)=0,\, D\Phi(g)\cdot \dot{g}=0\}\subset TG^r\end{equation} where $D\Phi(g)$ denotes the $(n\times \overline{m})$ Jacobian matrix of the contraints. 

Denote $\displaystyle{\mathfrak{g}^r:=\underbrace{\mathfrak{g}\times\ldots\times\mathfrak{g}}_{r-times}}$, where the Lie algebra structure of $\mathfrak{g}^r$ is given by $[\xi_1,\xi_2]=([\xi_1^1,\xi_2^1],\ldots,[\xi_1^r,\xi_2^r])\in\mathfrak{g}^r$ with $\xi_1=(\xi_1^1,\ldots,\xi_1^r)\in\mathfrak{g}^r$ and $\xi_2=(\xi_2^1,\ldots,\xi_2^r)\in\mathfrak{g}^r.$

We also denote $\tau_{i}:\mathfrak{g}^r\to\mathfrak{g}$ and $\beta_i:TG^r\to TG$  the corresponding canonical projections over its $i^{th}$-factors. Note that, after a left-trivialization, $T\mathcal{M}$ can be seen as a submanifold of $G^r\times\mathfrak{g}^r$ given by \begin{equation*}\label{tmtrivialized}\mathfrak{M}=\{(g,\xi)\in G^r\times\mathfrak{g}^r\mid\Phi(g)=0, \langle T^{*}_{\bar{e}}L_{g^{-1}}(D\Phi(g)),\xi\rangle=0\}\end{equation*}where $\xi=g^{-1}\dot{g}\in\mathfrak{g}^r$.

The control policy for the motion feasibility problem for multi-agent systems can be determined by solving the corresponding dynamics (trivialized Euler-Lagrange equations \eqref{eq_ep_intro}) for each $i\in\mathcal{N}$ subject to the constraint for each edge, as a unique system of differential equations, by lifting the dynamics of each vertex to $G^r\times\mathfrak{g}^r$, the constraints to $G^r$, and to study the dynamics for the formation problem as a holonomically constrained Lagrangian system on $G^r\times\mathfrak{g}^r$.

\subsection{Left-invariant kinematic multi-agent control systems}\label{kinematicsection}

Let $X:G^r\to TG^r$ be a vector field on $G^r$. The set $\mathfrak{X}(G^r)$ denotes the set of all vector fields on $G^r$. The tangent map $T_{\overline{e}}L_g$ shifts vectors based at $\overline{e}$ to vectors based at $g\in G^r$. By doing this operation for every $g\in G^r$ we define a vector field as $X_{\xi}^{g}:=T_{\overline{e}}L_g(\xi)$ for $\xi:=X(\overline{e})\in T_{\overline{e}}G^r$. 

\begin{definition}\label{defLI}A vector field $X\in\mathfrak{X}(G^r)$ is called \textit{left-invariant} if $T_{h}L_g(X(h))=X(L_g(h))=X(gh)$ $\forall$ $g,h\in G^r$.\end{definition}

 In particular for $h=\overline{e}$, Definition \ref{defLI} means that a vector field $X$ is left-invariant if $\dot{g}=X(g)=T_{\overline{e}}L_{g}\xi$ for $\xi=X(\overline{e})\in\mathfrak{g}^r$. Note that if $X$ is a left invariant vector field, then $\xi=X(\overline{e})=T_{g}L_{g^{-1}}\dot{g}$.

Consider an undirected static and connected graph $\mathcal{G}=(\mathcal{N},\mathcal{E}, \mathcal{C})$, describing the kinematics of each agent given by $r$ left invariant kinematic control systems each one on $G$, together with the constraints defining the set $\mathcal{C}$. As before, $\mathcal{N}$ denotes the set of vertices of the graph, but now, each $i\in\mathcal{N}$ is a fully-actuated left invariant kinematic control system, that is, the kinematics of each agent is determined by  \begin{equation}\label{eq:left-invariant-control-system-multi-agents}\dot{g}_i=T_{\overline{e}_i}L_{g_i}(\xi_i),\quad g_i(0)=g_0^{i},\end{equation} and the set $\mathcal{E}\subset\mathcal{N}\times\mathcal{N}$ denotes, as before, the set of edges for the graph, where $g_i(\cdot)\in C^{1}([a_i,b_i],G)$, $a_i,b_i\in\mathbb{R}^{+}$ is  fixed and $\xi_i$ is a curve on the Lie algebra $\mathfrak{g}$ of $G$. Alternatively, the left-invariant control system \eqref{eq:left-invariant-control-system-multi-agents} can be written as $\xi_i(t)=T_{g_i}L_{g_i^{-1}}\dot{g}_i$. Each curve $\xi_i(t)$ on the Lie algebra determines a control input $u_i(t)$, where for each $i$, the $n$-tuple of control inputs $u_i=[u^{1}_i\ldots u^{n}_i]^{T}$ takes values in $\mathbb{R}^{n}$.

If for each agent, $\mathfrak{g}= \hbox{span}\{e_{1},\ldots,e_{n}\}$, then $u_i$ satisfies $\displaystyle{
\xi_i(t) = \sum_{s=1}^{n}u^{s}_i(t)e_{s}}$ and therefore  \eqref{eq:left-invariant-control-system-multi-agents} is given by the \textit{drift-free kinematic left invariant control system} \begin{equation}\label{lics.1}\dot{g}_i=g_i\sum_{s=1}^{n}u^{s}_i(t)e_{s}.\end{equation}

Note that we have not made any reference to coordinates on $G^r$. We only require a basis for $\mathfrak{g}^r$. This is all that is necessary to study left-invariant kinematic systems.

\section{A differential-algebraic approach to characterize motion feasibility of LIMS}

In this section, inspired by  \cite{tabuada},  we consider a differential-algebraic approach for the motion feasibility problem for formation control of kinematic left-invariant multi-agent systems (LIMS's) introduced in Section \ref{kinematicsection}.

Given the collision avoidance constraint $\Phi:G^r\to\mathbb{R}$, consider the new constraint $d\Phi:TG^r\to\mathbb{R}$ and consider the corresponding projection into the $G\times G$ denoted by $\Phi_{ij}:G\times G\to\mathbb{R}$. 

\begin{definition} The constraint $d\Phi_{ij}(g)$ is said to be \textit{left invariant} if $T^{*}_{\overline{e}}L_{g^{-1}}d\Phi_{ij}(g)=d\Phi_{ij}(\bar{e})$, that is, the pullback to the identity of the constraint corresponds to the constraint at the identity.\end{definition}

Note that $d\Phi_{ij}\in TG^r$ and $d\Phi_{ij}$ evaluated at a point $g\in G^r$, i.e., $d\Phi_{ij}(g)$, is a one form on $G^r$.

 When the constraint is left-invariant, there exists a left-invariant distribution of feasible velocities for the formation given by the annhilator of the constraints at each point, and it  determines a subgroup of $G^r$. In other words, $\Phi^{-1}(0)\subset G^r$ is a subgroup of $G^r$ and classical reduction by symmetries \cite{Holmbook}, \cite{MaRa1999} can be performed in the multi-agent system to obtain an unconstrained reduced problem on $S=G^r/\Phi^{-1}(0)$. Given that the collision avoidance constraints in general are defined by the distances among configurations of the agents, we are mainly interested in constraints depending explicitly on the variables on $G^r$.

The co-Adjoint action on each Lie group $G$ induces a co-Adjoint action on $G^{r}$, denoted as  $\Ad^{*}_{g^{-1}}:(\mathfrak{g}^{*})^r\to(\mathfrak{g}^{*})^r$, and given by $\langle\Ad^{*}_{g^{-1}}\mu,\xi\rangle=\langle\mu,\Ad_{g^{-1}}\xi\rangle$ with $\mu\in(\mathfrak{g}^{*})^r$, $\xi\in\mathfrak{g}^r$ and $g^{-1}=(g^{-1}_1,\ldots, g^{-1}_r)$ the inverse element of $g\in G^{r}$.

To give a necessary condition for the existence of feasible motion in the formation problem for kinematic LIMS's, we want to find the set of  \textit{velocities satisfying the kinematics and the constraints, that is, the set of admissible velocities leaving the constraints invariant at a given point on $G^r$}. 

\begin{theorem}\label{theorem}
The set of admissible velocities allowing feasible motion in the formation problem is given by the set of elements $\xi\in\mathfrak{g}$ such that for a fixed $g\in G^r$, $$\langle \Ad_{g^{-1}}^{*}(T^{*}_{\overline{e}}L_{g^{-1}}(d\Phi_{ij}(g))),\xi\rangle=0.$$

\end{theorem}

\textit{Proof:}  
The interaction between agents in the formation, given by the formation constraints on $G\times G$ induces the constraint on $G^r$,  $\Phi_{ij}(g)=\phi_{ij}(\pi_{i}(g),\pi_j(g))$. Motion feasibility requires that the constraints holds along the trajectories of the LIMS \eqref{lics.1}. 

Differentiating the constraint $\Phi_{ij}(g)$ on $G^r$ we get the constraint $\dot{g}\in T_gG^r$, that is, $\langle d\Phi_{ij}(g),\dot{g}\rangle=0,$ where  $\dot{g}=[\dot{g}_1,\ldots,\dot{g}_{r}]^{T}$ with $\dot{g}_i\in T_{(\pi_{i}(g))}G$ and $d\Phi_{ij}(g)$ is a one-form on $G^r$, $d\Phi_{ij}(g)\in T^{*}_{g}G^r$.

 The one-forms $d\Phi_{ij}(g)\in T^{*}_{g}G^r$ can be translated back to $(\mathfrak{g}^{*})^r$ using left translations as $\langle d\Phi_{ij}(g),\dot{g}\rangle
=\langle T^{*}_{\overline{e}}L_{g^{-1}}(d\Phi_{ij}(g)),\xi\rangle$,
where $\xi=g^{-1}\dot{g}\in\mathfrak{g}^r$ and $ T^{*}_{\overline{e}}L_{g^{-1}}(d\Phi_{ij}(g))\in(\mathfrak{g}^{*})^r$ are the Lie algebra evaluated constraints, and where we used that $T(L_{g}\circ L_{{g^{-1}}})=TL_{g}\circ TL_{g^{-1}}$ is equal to the identity map on $TG^r$ and equation \eqref{relation-cltl}. 

Note that our problem not only involves that solutions must satisfy the constraints, which means that $\langle T^{*}_{\overline{e}}L_{g^{-1}}(d\Phi_{ij}(g)),\xi\rangle=0$ for $g\in G^r$, $g\neq\overline{e}$. Solutions must also be left invariant vector fields, solving \eqref{lics.1}, that is, $X(g)=gX(\overline{e})$ (see Definition \ref{defLI}). To solve the combined problem we proceed as in \cite{altafini2} (Section 4), to unify the solution in a unique algebraic condition. In order to find the left-invariant vector fields $X(g)$ satisfying the constraints, we must study how much the vector fields $X(g)\in T_{g}G^r$ changes from $\xi=X(\overline{e})\in\mathfrak{g}^r$. As a  transformation connecting $g$ and the identity $\overline{e}$ in $G^r$ we use the Adjoint operator, this means that we have to find $\xi\in\mathfrak{g}^r$ such that for a fixed $g\in G^r$ 
\begin{equation}\label{eq:formationcondition}\langle T^{*}_{\overline{e}}L_{g^{-1}}(d\Phi_{ij}(g)), \Ad_{g^{-1}}\xi\rangle=0.\end{equation} 

The operator $\Ad_{g^{-1}}$ represents a change of basis on $\mathfrak{g}^r$ and equation \eqref{eq:formationcondition} gives the subspace of $\mathfrak{g}^r$ annihilated by $T^{*}_{\overline{e}}L_{g^{-1}}(d\Phi_{ij}(g))$. Therefore the problem consists on finding $\xi\in\mathfrak{g}^{r}$ such that for a fixed $g\in G^r$, \begin{equation}\label{eq:formationcondition2}\langle \Ad_{g^{-1}}^{*}(T^{*}_{\overline{e}}L_{g^{-1}}(d\Phi_{ij}(g))),\xi\rangle=0.\end{equation}

\begin{remark}
As pointed out in \cite{altafini2}, equation \eqref{eq:formationcondition2} gives a linear system of algebraic equations with $\displaystyle{\xi=\sum_{i=1}^{n}u_ie_i}$ as unknown variables and $\Ad_{g^{-1}}^{*}(T^{*}_{\overline{e}}L_{g^{-1}}(d\Phi_{ij}(g)))$ as the known coefficients. Thus, solutions of the latter equation give rise to linear combinations of the control inputs in the linear subspace of $\mathfrak{g}$ annihilating the constraints.This means that in order for the solution not leave the submanifold which defines the constraints, the set of velocities must satisfy equations \eqref{eq:formationcondition2} 
\end{remark}

\begin{remark}
Note that the one-forms $d\Phi_{ij}(g)$ are not left-invariant and may change at any point $g\in G^r$, but by using the co-adjoint action, we can study vector fields at any point $g\in G^r$. Therefore, the problem of finding the orthogonal subspace of the constraints at points of $G^r$ given in \cite{tabuada}, to characterize the physical allowable directions of motion, in the context of LIMS's, is equivalent to finding the annihilator of the co-adjoint action for $T^{*}_{\overline{e}}L_{g^{-1}}(d\Phi_{ij}(g))$.
\end{remark}
\begin{remark}\label{abstraction} Following \cite{tabuada} (Section III-B), when more than one solutions exist, the solution space can be exploited to find a new distribution which is called in \cite{tabuada} a group \textit{abstraction} of the kinematic LIMS \eqref{lics.1}, that is, a new control system keeping the formation along solutions. This new control system is given by studying the kernel of the co-distribution defined by the union of a basis of $d\Phi$ and a basis for the co-distribution describing the kinematics of the agents.

For LIMS's, the space of solutions $\mathcal{O}\subset\mathfrak{g}^r$ is determined by equation \eqref{eq:formationcondition2}, that is, $$\mathcal{O}:=\{\xi\in\mathfrak{g}^r|\langle \Ad_{g^{-1}}^{*}(T^{*}_{\overline{e}}L_{g^{-1}}(d\Phi_{ij}(g))),\xi\rangle=0\}$$ for a fixed $g\in G^r$. As in \cite{tabuada}, one may use $\mathcal{O}$ to find an abstraction for LIMS. The new control system is given by the Kernel of $\mathcal{O}$, that is, $$G_{\mathcal{O}}:=\{\eta\in\mathcal{O}|\langle \Ad_{g^{-1}}^{*}(T^{*}_{\overline{e}}L_{g^{-1}}(d\Phi_{ij}(g))),\eta\rangle=0\}=\hbox{Ker}(\mathcal{O})$$ giving rise to a group abstraction that describes the set of admissible velocities keeping the formation of the LIMS.

By considering a basis of $G_{\mathcal{O}}$, denoted $\{K_1,\ldots,K_s\}$, we can write such a group abstraction as the left-invariant control system (note that $\hbox{Ker}(\mathcal{O})$ does not depend on $G$ and therefore its basis is given by left-invariant elements) \begin{equation}\label{abstracted}\dot{g}=\sum_{k=1}^{s}K_{k}\omega_k\end{equation}
where $\omega_k$ are the new control inputs that activate the elements of the base $\{K_1,\ldots,K_s\}$ with $s\leq \dim(G_{\mathcal{O}})$. The abstracted control system \eqref{abstracted} provides certain insights on different types of feasible motions for the agent according to different choices of $\omega_k$ for $k=1,\ldots,s$. 

As pointed out in \cite{tabuada}, since for all $k=1,\ldots,s$, $\langle \Ad_{g^{-1}}^{*}(T^{*}_{\overline{e}}L_{g^{-1}}(d\Phi_{ij}(g))),K_{k}\rangle=0$, all inputs $\omega_k$ gives rise to trajectories $g(t)$ satisfying the left invariant multi-agent control system \eqref{eq:left-invariant-control-system-multi-agents} and the formation constraints $\Phi_{ij}(g)$.\end{remark}

\begin{example}\label{example1} As an application we consider the motion feasibility problem for three agents moving in the plane. The configuration of each agent at any given time is determined by the element $g_i\in\mathrm{SE}(2)\cong\mathbb{R}^{2}\times\mathrm{SO}(2)$, $i=1,2,3$ given by $\displaystyle{g_i = \begin{bmatrix}
    \cos\theta_i & -\sin\theta_i & \phantom{-}x_i\\
    \sin\theta_i & \phantom{-}\cos\theta_i & \phantom{-}y_i\\
    0 & \phantom{-}0 & \phantom{-}1
  \end{bmatrix},\nonumber}$ where $p_i=(x_i, y_i)\in\mathbb{R}^{2}$.

The kinematic equations for the multi-agent system are
\begin{equation}\label{unicycle}
\dot{p}_i = R_iu_i,\quad 
\dot{R}_i= R_iu_i, \quad i=1,2,3, \hbox{with } u_i=(u_i^{1},u_i^{2})
\end{equation} where $R_i=\left(
     \begin{array}{cc}
      \cos\theta_i& -\sin\theta_i\\
       \sin\theta_i &\cos\theta_i \\
     \end{array}
   \right)\in SO(2).$

The Lie algebra $\mathfrak{se}(2)$ of $SE(2)$ is determined by
\begin{equation}\label{se2def}\mathfrak{se}(2)=\Big{\{}\left(
     \begin{array}{cc}
      A & b\\
       0 & 0 \\
     \end{array}
   \right): A\in \mathfrak{so}(2) \hbox{ and } b\in \mathbb{R}^2\Big{\}}\end{equation} where $A=-aJ$, $a\in\mathbb{R}$, with $J=\left(
     \begin{array}{cc}
      0& 1\\
       -1 &0 \\
     \end{array}
   \right)$ and we identify the Lie algebra $\mathfrak{se}(2)$ with $\mathbb{R}^3$ via the isomorphism
   $\displaystyle{\left(
     \begin{array}{cc}
     -aJ& b\\
       0 & 0 \\
     \end{array}
   \right)\mapsto (a,b)}$.

Equations \eqref{unicycle} gives rise to a left-invariant control system on $(SE(2))^3$ with the form $\dot{g}_i=g_i(e_1^{i}u_i^{1}+e_2^{i}u_i^{2})$ describing all directions of allowable motion, where the elements of the basis of $\mathfrak{g}=\mathfrak{se}(2)$ are 

\[\begin{small}
e_{1}^i = \begin{bmatrix}
        0 & \phantom{-}0 & \phantom{-}1\\
        0 & \phantom{-}0 & \phantom{-}0\\
        0 & \phantom{-}0 & \phantom{-}0
        \end{bmatrix}, e^i_{2} = \begin{bmatrix}
        0 & \phantom{-}0 & \phantom{-}0\\
        0 & \phantom{-}0 & \phantom{-}1\\
        0 & \phantom{-}0 & \phantom{-}0
        \end{bmatrix}, e_{3}^i = \begin{bmatrix}
        0 & -1 & \phantom{-}0\\
        1 & \phantom{-}0 & \phantom{-}0\\
        0 & \phantom{-}0 & \phantom{-}0
        \end{bmatrix},\end{small}\]which satisfy
$[e_{3}^i,e_{2}^i] = e_{1}^i, \hspace{5pt} [e_{1}^i,e_{2}^i] = 0_{3\times 3}, \hspace{5pt} [e_{1}^i,e_{3}^i] = e_{2}^i$.
Using the dual pairing, where $\langle\alpha_i,\xi_i\rangle:=\hbox{tr}(\alpha_i\xi_i)$, for any $\xi_i\in\mathfrak{se}(2)$ and $\alpha_i\in\mathfrak{se}(2)^{*}$, the elements of the basis of $\mathfrak{se}(2)^{*}$ are given by \[\begin{small}
 e^{1}_i = \begin{bmatrix}
        0 & \phantom{-}0 & \phantom{-}0\\
        0 & \phantom{-}0 & \phantom{-}0\\
        1 & \phantom{-}0 & \phantom{-}0
        \end{bmatrix}, e^{2}_i = \begin{bmatrix}
        0 & \phantom{-}0 & \phantom{-}0\\
        0 & \phantom{-}0 & \phantom{-}0\\
        0 & \phantom{-}1 & \phantom{-}0
        \end{bmatrix},\end{small}\]
       \[\begin{small}  e^{3}_i = \begin{bmatrix}
        \phantom{-}0 & \phantom{-}\frac{1}{2} & \phantom{-}0\\
        -\frac{1}{2} & \phantom{-}0 & \phantom{-}0\\[8pt]
        \phantom{-}0 & \phantom{-}0 & \phantom{-}0
        \end{bmatrix}.\end{small}\]
The communication topology is given by an equilateral triangle where each node communicates with its adjacent vertex.
 
The formation is completely specified by the (holonomic) constraints $\phi_{ij}^{k}:SE(2)\times SE(2)\to\mathbb{R}$, (i.e., $\phi_{12}^{1},\phi_{13}^{2},\phi_{23}^{3}$) determined by a prescribed distance $d_{ij}$ among the positions of all agent at any time. The constraint for the edge $e_{ij}$ is given by  $\phi_{ij}(g_i,g_j)=||\psi(g_j)g_i||^2_{F}-\tilde{d}_{ij}$ where $||\cdot||_{F}$ is the Frobenius norm, $||A||_{F}=\hbox{tr}\left(A^{T}A\right)^{1/2}$, $\tilde{d}_{ij}=d_{ij}^2+3$ and $\psi:SE(2)\to SE(2)$ is the smooth map defined as $\psi(g)=\bar{g}$ where $\bar{g}=\left[ {\begin{array}{ccc}
   1 & 0 & -x \\
   0 & 1 & -y \\
   0 & 0 &1\\
  \end{array} } \right]\in SE(2).$ 
    
  It is straightforward to check that the constraint $\phi_{ij}^{k}(g_i,g_j)=0$ on absolute configurations on the Lie group $SE(2)\times SE(2)$, is equivalent to the constraint in the relative configurations, that is, $\phi_{ij}^{k}(g_i,g_j)=0$ is equivalent to $(x_i-x_j)^2+(y_i-y_j)^2-d_{ij}^2=0$.

The inner product on $\mathfrak{se}(2)$ is given by $\langle\langle\xi_i,\xi_i\rangle\rangle_{\mathfrak{se}(2)}=\hbox{tr}(\xi_i^{T}\xi_i)$, for any $\xi_i\in\mathfrak{se}(2)$ and hence, the norm $\|\xi_i\|_{\mathfrak{se}(2)}$ is given by $\|\xi_i\|_{\mathfrak{se}(2)}=\langle\langle\xi_i,\xi_i\rangle\rangle^{1/2}_{\mathfrak{se}(2)} = \sqrt{\hbox{tr}(\xi_i^{T}\xi_i)}$, $\xi_i\in\mathfrak{se}(2)$.

Equations \eqref{unicycle} are a set of equations on the Lie algebra $\mathfrak{se}(2)\times\mathfrak{se}(2)\times\mathfrak{se}(2)$ which together with the set of constraints $\mathcal{C}=\{\phi_{12}^{1}, \phi_{13}^{2}, \phi_{23}^{3}\}$ specify the formation for the multi-agent control system. 

To apply Theorem \ref{theorem} we need the expression for the co-Adjoint action on $\mathfrak{se}(2)^{*}$. The co-Adjoint action of $SE(2)$ on $\mathfrak{se}(2)^{*}$, denoted $\Ad^{*}:SE(2)\times\mathfrak{se}(2)^{*}\to\mathfrak{se}(2)^{*}$ is given by \begin{equation}\label{adestrella}\Ad_{(R_{i},p_i)^{-1}}^{*}(\mu_i,\beta_i)=(\mu_i-R_{i}\beta_i\cdot Jp_i, R_{i}p_i)\in\mathfrak{se}(2)^{*},\end{equation} where we are using the notation $g_i=(R_{i},p_i)\in SO(2)\times\mathbb{R}^{2}=SE(2)$, $\mu_i\in\mathfrak{se}(2)^{*}$, $\beta_i\in\mathbb{R}^{2}$ and $J$ as in \eqref{se2def}.

Denote $g_{ij}:=\psi(g_j)^{T}\psi(g_j)g_i\in SE(2)$ and $\bar{g}_{ij}:=\psi(g_j)g_ig_i^{T}\in SE(2)$. The matrix $T^{*}_{\overline{e}}L_{g^{-1}}(d\Phi_{ij}(g)))$ in terms of the basis for $\mathfrak{se}(2)^{*}\times\mathfrak{se}(2)^{*}\times\mathfrak{se}(2)^{*}$ is given by  \begin{align}
&T^{*}_{\overline{e}}L_{g^{-1}}(d\Phi_{ij}(g))= \begin{bmatrix}
         T^{*}_{\overline{e}_1}L_{g_1^{-1}}(d\phi_{12}^{1})(g_1,g_2)\\
        T^{*}_{\overline{e}_1}L_{g_1^{-1}}(d\phi_{13}^{2})(g_1,g_3)\\
        T^{*}_{\overline{e}_2}L_{g_2^{-1}}(d\phi_{23}^{3})(g_2,g_3)
        \end{bmatrix}\label{Tphi}\\\
        &=\begin{bmatrix}
        (2g_{12})_{31}e_1^1+(2g_{12})_{32}e_{1}^{2}+(2\bar{g}_{12})_{31}e_2^1+(2\bar{g}_{12})_{32}e_{2}^{2}\\
         (2g_{13})_{31}e_1^1+(2g_{13})_{32}e_{1}^{2}+(2\bar{g}_{13})_{31}e_3^1+(2\bar{g}_{13})_{32}e_{3}^{2}\\
         (2g_{23})_{31}e_2^1+(2g_{23})_{32}e_{2}^{2}+(2\bar{g}_{23})_{31}e_3^1+(2\bar{g}_{23})_{32}e_{3}^{2}
        \end{bmatrix}\nonumber\end{align}where the subindexes $13$ and $23$ stands for the entry $31$ and $32$ of the matrices $g_{ij}$ and $\bar{g}_{ij}$, where we have used that $\frac{d}{dX}\hbox{tr}(X^{T}BX)=BX+B^{T}X$ for matrices $X$ and $B$, to compute \begin{align*}\frac{\partial\phi_{ij}}{\partial g_i}=&\frac{\partial}{\partial g_i}\hbox{tr}(g_i ^{T}\psi(g_j)^{T}\psi(g_j)g_i)\\=&\psi(g_j)^{T}\psi(g_j)g_i+\psi(g_j)^{T}\psi(g_j)g_i=2\psi(g_j)^{T}\psi(g_j)g_i\end{align*} and similarly we used $\frac{\partial}{\partial X}\hbox{tr}(B^TX^TXA)=2XAA^{T}$, to obtain $\displaystyle{\frac{\partial\phi_{ij}}{\partial g_j}=2\psi(g_j)g_ig_i^{T}}$. Combining \eqref{adestrella} and \eqref{Tphi}, by Theorem \ref{theorem}, there are trajectories for each agent  satisfying the formation constraints as well the kinematics given by left-invariant vector fields.

\end{example}

\section{Variational characterization for formation control of multi-agent systems on Lie groups}\label{section5}
In this section we study the motion of dynamical multi-agent control systems on a Lie group by applying techniques from variational calculus, after a left-trivialization of the tangent bundle. In order to determine the dynamics for the formation problem we use the Lagrange multipliers Theorem \ref{LMth}.

Assume that the dynamics of each agent is described by a Lagrangian function $\ell_i:G\times\mathfrak{g}\to\mathbb{R}$ and define the overall Lagrangian function $\mathbf{L}:G^r\times\mathfrak{g}^r\to\mathbb{R}$ by \begin{equation}\label{lagrangianL}\mathbf{L}(g,\xi)=\sum_{i=1}^{r}\ell_i(\pi_i(g),\tau_i(\xi))\end{equation} with $\tau_{i}:\mathfrak{g}^r\to\mathfrak{g}$ defined as in Section \ref{lim}.

In the variational principle developed below we introduce the formation constraints into the dynamics by incorporating the factor $\displaystyle{\frac{1}{2}\sum_{j\in\mathcal{N}_i}\sum_{k=1}^{\overline{m}}\lambda_k\Phi_{ij}^{k}(g)}$ into \eqref{lagrangianL}, with $\lambda_k\in\mathbb{R}$ being the Lagrange multipliers and $\overline{m}$ as in \eqref{mbar}. The factor $\frac{1}{2}$ in the previous summation is done in order to not count twice the quantity of functions $\Phi_{ij}^k$ (note that $\Phi_{ij}^k=\Phi_{ji}^k$).
This approach permits to study the formation problem from a  decentralized perspective (see for instance \cite{MMbook} Section $6.5.2$).

Let us denote by $\mathcal{C}(G^r\times\mathfrak{g}^r)=\mathcal{C}([0,T], G^r\times\mathfrak{g}^r, g_0, g_T)$ the space of smooth functions $(g,\xi):[0,T]\to G^r\times\mathfrak{g}^r$ satisfying $g(0)=g_0$, $g(T)=g_T$. Denote also by $\mathcal{C}(\mathbb{R}^{\overline{m}})=\mathcal{C}([0,T],\mathbb{R}^{\overline{m}})$ the space of curves $\lambda:[0,T]\to\mathbb{R}^{\overline{m}}$ in $\mathbb{R}^{\overline{m}}$,  without boundary conditions.  

 The action functional $\mathcal{S}^{G^r\times\mathfrak{g}^r}:\mathcal{C}(G^r\times\mathfrak{g}^r)\to\mathbb{R}$ for $\mathbf{L}:G^r\times\mathfrak{g}^r\to\mathbb{R}$ is given by 
$
\displaystyle{\mathcal{S}^{G^r\times\mathfrak{g}^r}(g,\xi)=\int_{0}^{T}\mathbf{L}(g,\xi)\,dt.}$ 

Consider the augmented Lagrangian $\overline{L}:G^r\times\mathfrak{g}^r\times\mathbb{R}^{\overline{m}}\to\mathbb{R}$ given by $\displaystyle{\overline{L}(g,\xi,\lambda)=\mathbf{L}(g,\xi)-\lambda\cdot\Phi(g)},$ with $\cdot$ being the dot product on $\mathbb{R}^{\overline{m}}$. Note that such an extended Lagrangian can be associated with an action functional $\overline{\mathcal{S}}:\mathcal{C}(G^r\times\mathfrak{g}^r\times\mathbb{R}^{\overline{m}})\to\mathbb{R}$ given by $$\overline{\mathcal{S}}(g,\xi,\lambda):=\int_{0}^{T}\mathbf{L}(g,\xi)\,dt-\langle\langle\lambda,\Phi(g)\rangle\rangle$$ where $\langle\langle\cdot,\cdot\rangle\rangle$ denotes the $L^2$ inner product \footnote{recall that given two functions from $[0,T]$ to $\mathbb{R}^{\overline{m}}$,  $\langle\langle f,g\rangle\rangle=\int_{0}^{T}f\cdot g\,dx$} on $\mathbb{R}^{\overline{m}}$. 

To prove Theorem \ref{theorem1var} below we need to introduce the class of infinitesimal variations we shall consider in the variational principle.
\begin{definition} Let $g:[0,T]\to\mathbb{R}$ be a curve on $G$. For $\epsilon>0$, the \textit{variation} of the curve $g$ is the family of differentiable curves on $G$, $g_{\epsilon}:(-\epsilon,\epsilon)\times [0,T]\to G$ such that $g_0(t)=g(t)$. The \textit{infinitesimal variation} of $g$ is defined by $\delta g=\frac{d}{d\epsilon}g_{\epsilon}(t)\mid_{\epsilon=0}$.\end{definition}

\begin{remark}\label{eta}
Given that $\xi=g^{-1}\dot{g}$, infinitesimal variations for $\xi$ are induced by infinitesimal variations of $g$, that is $\delta\xi=\dot{\eta}+ad_{\xi}\eta$ where $\eta$ is an arbitrary path in $\mathfrak{g}$ defined by $\eta=T_{g}L_{g^{-1}}(\delta g)=g^{-1}\delta g$, that is $\delta g=g\eta$ (see \cite{Holmbook} Section $7.3$, p. 255 for the proof) and where from the last equality is follows that variations of $g$ vanishing at the end points implies that $\eta$ must vanish at end points, that is, $\eta(0)=\eta(T)=0$ since $g(0)=g_0$ and $g(T)=g_T$ are not necessarily zero.\end{remark}\vspace{.1cm}

\begin{theorem}\label{theorem1var}  If $(g,\xi)\in\mathcal{C}(G^r\times\mathfrak{g}^r)$ is an extrema of  $\mathcal{S}^{G\times\mathfrak{g}}$, and hence solves the Euler-Lagrange equations for $\mathbf{L}$, then $(g,\xi,\lambda)\in\mathcal{C}(G^r\times\mathfrak{g}^r\times\mathbb{R}^{\overline{m}})$ is an extrema of $\overline{\mathcal{S}}$ and hence solves the constrained Euler-Lagrange equations for the augmented Lagrangian $\overline{L}$ given by 
 \begin{align*}
0=&\frac{d}{dt}\left(\frac{\partial \ell_i}{\partial \xi_i}\right)-\hbox{ad}^{*}_{\xi_i}\left(\frac{\partial \ell_i}{\partial \xi_i}\right)+T^{*}_{\overline{e}_i}L_{g_i}\left(\frac{\partial \ell_{i}}{\partial g_i}\right)\\
&-\sum_{j\in\mathcal{N}_i}\sum_{k=1}^{\overline{m}}\lambda_k\left(T_{\overline{e}_{i}}^{*}L_{g_i}\frac{\partial \phi_{ij}^k}{\partial g_i}\right),\quad i=1,\ldots,r\nonumber\\
0=&\phi_{ij}^{k}(g_i,g_j) \hbox{ for all }k=1,\ldots,\overline{m},\quad i=1,\ldots,r,\quad j\in\mathcal{N}_i.\label{eqCEL2}
\end{align*}\end{theorem}

\textit{Proof:} If $(g,\xi)\in T\mathcal{M}$ is an extrema of $\mathcal{S}^{G^r\times\mathfrak{g}^r}$, then by the Lagrange multiplier theorem, $(g,\xi,\lambda)\in\mathcal{C}(G^r\times\mathfrak{g}^r)\times\mathcal{C}(\mathbb{R}^{\overline{m}})$ is an extrema of $\overline{\mathcal{S}}(g,\xi,\lambda)=\mathcal{S}^{G^r\times\mathfrak{g}^r}(g,\xi)-\langle\langle\lambda,\Phi(g)\rangle\rangle$. By identifying $\mathcal{C}(G^r\times\mathfrak{g}^r)\times\mathcal{C}(\mathbb{R}^{\overline{m}})$ with $\mathcal{C}(G^r\times\mathfrak{g}^r\times\mathbb{R}^{\overline{m}})$ we note that  
\begin{align*}
\overline{\mathcal{S}}(g,\xi,\lambda)=&\mathcal{S}^{G^r\times\mathfrak{g}^r}(g,\xi)-\langle\langle\lambda,\Phi(g)\rangle\rangle\\
=&\int_{0}^{T}\left(\mathbf{L}(g,\xi)-\lambda\cdot\Phi(g)\right)\,dt
\end{align*} is the action functional for the augmented Lagrangian $\overline{L}(g,\xi,\lambda)=\mathbf{L}(g,\xi)-\lambda\cdot\Phi(g)$, where we have used the definition of $\langle\langle\cdot,\cdot\rangle\rangle$. As $(g,\xi,\lambda)\in\mathcal{C}(G^r\times\mathfrak{g}^r\times\mathbb{R}^{\overline{m}})$ must extremize this action, it is a solution of the Euler-Lagrange equations for $\overline{L}$. 

Next, we extremizes $\overline{\mathcal{S}}$ by solving $d\overline{\mathcal{S}}=0$ to obtain the Euler-Lagrange equations for $\overline{L}$. The action integral $\overline{\mathcal{S}}$ along a variation of the motion is $\displaystyle{\overline{\mathcal{S}}_{\epsilon}=\int_{0}^{T}\overline{L}(g_{\epsilon},\xi_{\epsilon},\lambda_{\epsilon})\,dt.}$ The varied value of this action functional can be expressed as a power series in $\epsilon$, that is, $\overline{\mathcal{S}}_{\epsilon}=\overline{\mathcal{S}}+\epsilon\delta\overline{\mathcal{S}}+\mathcal{O}(\epsilon^2)$ where the infinitesimal variation of $\overline{\mathcal{S}}$ is given by $\displaystyle{\delta\overline{\mathcal{S}}=\frac{d}{d\epsilon}\overline{S}_{\epsilon}\Big{|}_{\epsilon=0}}$. 

Hamilton's principle states that the infinitesimal variation of $\overline{\mathcal{S}}$ along any motion must be zero, that is, $\delta\overline{\mathcal{S}}=0$ for all possible infinitesimal variations in $(G^r\times\mathfrak{g}^r\times\mathbb{R}^{\overline{m}})$, where infinitesimal variations on $\mathfrak{g}$ are given by curves $\eta:[0,T]\to\mathfrak{g}$ satisfying $\eta(0)=\eta(T)=0$ (see Remark \ref{eta}).

Now, note that 
\begin{align*}
&\delta\int_{0}^{T}\overline{L}(g,\xi,\lambda)dt=\delta\int_{0}^{T}\left(\sum_{i=1}^{r}\ell_i(g_i,\xi_i)\right.\\&\qquad\qquad\qquad\qquad\qquad\left.-\frac{1}{2}\sum_{j\in\mathcal{N}_i}\sum_{k=1}^{\overline{m}}\lambda_k\phi_{ij}^{k}(g_i,g_j)\,\right)dt\\
\\=&\int_{0}^{T}\Big{\langle}\frac{\partial \ell_i}{\partial \xi_i},\delta \xi_i\Big{\rangle}-\frac{1}{2}\sum_{j\in\mathcal{N}_i}\sum_{k=1}^{\overline{m}}\phi_{ij}^{k}(g_i,g_j)\delta\lambda_k+\Big{\langle}\frac{\partial \ell_i}{\partial g_i},\delta g_i\Big{\rangle}\\
&-\frac{1}{2}\sum_{j\in\mathcal{N}_i}\sum_{k=1}^{\overline{m}}\lambda_k\left(\Big{\langle}\frac{\partial \phi_{ij}^k}{\partial g_i},\delta g_i\Big{\rangle}+\Big{\langle}\frac{\partial \phi_{ij}^k}{\partial g_j},\delta g_j\Big{\rangle}\right)\,dt\\
=&\int_{0}^{T}\Big{\langle}\frac{\partial \ell_i}{\partial \xi_i},\dot{\eta}_{i}+\hbox{ad}_{\xi_i}\eta_{i}\Big{\rangle}\\
&-\frac{1}{2}\sum_{j\in\mathcal{N}_i}\sum_{k=1}^{\overline{m}}\phi_{ij}^{k}(g_i,g_j)\delta\lambda_k+\Big{\langle}\frac{\partial \ell_i}{\partial g_i},\delta g_i\Big{\rangle}\\
&-\frac{1}{2}\sum_{j\in\mathcal{N}_i}\sum_{k=1}^{\overline{m}}\lambda_k\left(\Big{\langle}\frac{\partial \phi_{ij}^k}{\partial g_i},\delta g_i\Big{\rangle}+\Big{\langle}\frac{\partial \phi_{ij}^k}{\partial g_j},\delta g_j\Big{\rangle}\right)\,dt\end{align*}where from the second equality the sum over $i=1,\ldots,r$ has been omitted to reduce space,  and in third equality we replaced the variations on $\xi_i$ by their corresponding expressions (see Remark \ref{eta}). 

The first component of the previous integrand, after applying integration by parts twice, using the boundary conditions for $\eta_i$ and the definition of co-adjoint action, results in \begin{equation*}\label{integrand1}
\int_{0}^{T}\Big{\langle}-\frac{d}{dt}\left(\frac{\partial \ell_i}{\partial \xi_i}\right)+\hbox{ad}^{*}_{\xi_i}\left(\frac{\partial \ell_i}{\partial \xi_i}\right),\eta_i\Big{\rangle}.
\end{equation*} Using the fact that $T(L_{g_i}\circ L_{{g_i^{-1}}})=TL_{g_i}\circ TL_{g_i^{-1}}$ is equal to the identity map on $TG$ and $\eta_i=T_{g_i}L_{g_i^{-1}}(\delta g_i)$ (see Remark \ref{eta}), the third component can be written as \begin{equation*}\Big{\langle}\frac{\partial\ell_i}{\partial g_i},\delta g_i\Big{\rangle}=\Big{\langle}T^{*}_{\overline{e}_i}L_{g_i}\left(\frac{\partial\ell_i}{\partial g_i}\right),\eta_i\Big{\rangle}\end{equation*}

For the last member of the integrand we observe the following,\begin{align*}
&\frac{1}{2}\sum_{j\in\mathcal{N}_i}\sum_{k=1}^{\overline{m}}\lambda_k\left(\Big{\langle}\frac{\partial \phi_{ij}^k}{\partial g_i},\delta g_i\Big{\rangle}+\Big{\langle}\frac{\partial \phi_{ij}^k}{\partial g_j},\delta g_j\Big{\rangle}\right)=\\
&\frac{1}{2}\sum_{j\in\mathcal{N}_i}\sum_{k=1}^{\overline{m}}\lambda_k\left(\Big{\langle}T_{\overline{e}_{i}}^{*}L_{g_i}\frac{\partial \phi_{ij}^k}{\partial g_i},\eta_i\Big{\rangle}+\Big{\langle}T_{\overline{e}_j}^{*}L_{g_j}\frac{\partial \phi_{ij}^k}{\partial g_j},\eta_j\Big{\rangle}\right),
\end{align*}where we used the definition of left action and Eq. \eqref{relation-cltl}. Using the fact that $\phi_{ij}^{k}=\phi_{ji}^{k}$ in second term of the last expression, last sum can be written as \begin{equation}\label{eqq3}\frac{1}{2}\sum_{j\in\mathcal{N}_i}\sum_{k=1}^{\overline{m}}\lambda_k\left(\Big{\langle}T_{\overline{e}_{i}}^{*}L_{g_i}\frac{\partial \phi_{ij}^k}{\partial g_i},\eta_i\Big{\rangle}+\Big{\langle}T_{\overline{e}_{j}}^{*}L_{g_j}\frac{\partial \phi_{ji}^k}{\partial g_j},\eta_j\Big{\rangle}\right).\end{equation}

By employing a change of variables in second factor of the last expression, \eqref{eqq3} can be written as $\displaystyle{\sum_{j\in\mathcal{N}_i}\sum_{k=1}^{\overline{m}}\lambda_k\Big{\langle}T_{\overline{e}_{i}}^{*}L_{g_i}\frac{\partial \phi_{ij}^k}{\partial g_i},\eta_i\Big{\rangle}}$.

Therefore, $\displaystyle{\delta\int_{0}^{T}\overline{L}(g(t),\xi(t),\lambda(t))dt=0}$, for all $\delta\eta_{i}$, $\delta g_i$ and $
\delta\lambda_k$ implies 

 \begin{align}
0=&\frac{d}{dt}\left(\frac{\partial \ell_i}{\partial \xi_i}\right)-\hbox{ad}^{*}_{\xi_i}\left(\frac{\partial \ell_i}{\partial \xi_i}\right)+T^{*}_{\overline{e}_i}L_{g_i}\left(\frac{\partial \ell_{i}}{\partial g_i}\right)\label{eqCEL1}\\
&-\sum_{j\in\mathcal{N}_i}\sum_{k=1}^{\overline{m}}\lambda_k\left(T_{\overline{e}_{i}}^{*}L_{g_i}\frac{\partial \phi_{ij}^k}{\partial g_i}\right),\nonumber\\
0=&\phi_{ij}^{k}(g_i,g_j).\label{eqCEL2}
\end{align}
  
 Finally to describe the dynamics into the Lie group and therefore obtain the absolute configurations $g(t)\in G^r$ we must also consider the kinematics equation \begin{equation}\label{kinematicsCEL3}\dot{g}_i=g_i\xi_i\end{equation} with values in $\mathfrak{g}$, for each $i=1,\ldots,r$. Hence, the constrained Euler-Lagrange equations \eqref{eqCEL1}-\eqref{eqCEL2} and the kinematic equation \eqref{kinematicsCEL3} defines the Lagrangian flow on $TG^r\times\mathbb{R}^{\overline{m}}$ described by $(g,\dot{g},\lambda)\in TG^r\times\mathbb{R}^{\overline{m}}$, and therefore the set of differential equations \eqref{eqCEL1}-\eqref{kinematicsCEL3} gives rise to necessary conditions for the existence of feasible motion in the multi-agent system under collision avoidance constraints.\hfill\hfil$\square$

\begin{remark}\label{feedback}
Note that a feedback control from the motion feasibility problem can be constructed by solving the left-trivialized constrained Euler-Lagrange equations \eqref{eqCEL1}-\eqref{kinematicsCEL3} and using the solution to construct the feedback law $u_i$ employing equation \eqref{eq_ep_intro}. The existence of solutions for the equations of motion is guaranteed under a regularity condition as follows (see \cite{arnold} Section $1.4.2$): if the matrix \begin{equation}\label{regularity}\left(
     \begin{array}{cc}
      \frac{\partial^2\ell}{\partial\xi_i\partial\xi_i} & T^{*}_{\bar{e}_i}L_{g_i}\left(\frac{\partial\phi_{ij}^{k}}{\partial g_i}\right)\\
     \left(T^{*}_{\bar{e}_i}L_{g_i}\left(\frac{\partial\phi_{ij}^{k}}{\partial g_i}\right) \right)^{T}  & 0 \\
     \end{array}
   \right)\end{equation} is non-singular at every point in an open neighborhood $\mathcal{U}$ of the vector space $\mathfrak{g}^{r}\times\mathbb{R}^{\overline{m}}$ then there exists a unique solution $\gamma(t):=(g(t),\xi(t))\in G^{r}\times\mathfrak{g}^{r}$ of the Euler-Lagrange equations for $\ell$ with boundary values $\gamma(0)=\gamma_0$ and $\gamma(T)=\gamma_1$ with $\gamma_0,\gamma_1\in \mathcal{U}$ and satisfying the collision avoidance constraints.\end{remark}

\section{Application to multiple underwater vehicles}\label{sectionexample}
\subsection{System model}
Consider the collision avoidance problem for three rigid bodies evolving on the special Euclidean group $SE(3)$. Any element of $SE(3)$ is given by $\displaystyle{g_i=\begin{bmatrix} 
R_i & b_i \\
0 & 1
\end{bmatrix}}$ with $R_i\in SO(3)$ describing the orientation for the $i^{th}$-body as a rotation matrix and $b_i=(b_i^{x},b_i^{y},b_i^{z})\in\mathbb{R}^{3}$ is the position of the center of mass for the $i^{th}$-body in the inertial frame of coordinates.

For the shake of simplicity we write $g_i=(b_i,R_i)\in SE(3)\simeq\mathbb{R}^{3}\times SO(3)$. Therefore the state of each agent evolves in the $12$ dimensional tangent bundle $TSE(3)$. This space can be left-trivialized as $TSE(3)\simeq SE(3)\times\mathfrak{se}(3)$, where $\mathfrak{se}(3)\simeq\mathfrak{so}(3)\times\mathbb{R}^{3}\simeq\mathbb{R}^{3}\times\mathbb{R}^{3}$, with $\mathfrak{so}(3)$ denoting the space of $(3\times 3)$-skew-symmetric matrices. 
We denote by $\hat{\cdot}:\mathbb{R}^{3}\to\mathfrak{so}(3)$ the isomorphism between vectors on $\mathbb{R}^{3}$ and skew-symmetric matrices, given by  $$\hat{\Omega}_i(t)=\left(
  \begin{array}{ccc}
    0& -\Omega^3_i(t) & \Omega^2_i(t) \\
    \Omega^3_i(t) & 0 & -\Omega^1_i(t) \\
    -\Omega^2_i(t) & \Omega^1_i(t) & 0 \\
  \end{array}
\right)$$with $\Omega_i=(\Omega_i^{1},\Omega_{i}^{2},\Omega_i^{3})\in\mathbb{R}^{3}$.
The space $\mathfrak{se}(3)$ has elements $\displaystyle{\eta_i=\begin{bmatrix} 
\hat{\Omega}_i& \nu_i \\
0 & 1
\end{bmatrix}}$ where $\hat{\Omega}_i\in\mathfrak{so}(3)$, $\nu_i\in\mathbb{R}^{3}$. Using the inverse map of the isomorphism $\hat{\cdot}$, $\eta_i$ can be identified with the element $(\nu_i,\Omega_i)\in\mathbb{R}^{6}$, where $\nu_i$ is the translational velocity and $\Omega_i$ the angular velocity for the $i^{th}$ agent, both  in body coordinates. For the remainder of the paper, we represent the attitude state as an element of $SE(3)\times\mathfrak{se}(3)\simeq SE(3)\times\mathbb{R}^{6}$.

The kinematic equations are given by \begin{equation}\label{kinematicSE(3)}
\dot{R}_i=R_i\hat{\Omega}_i,\quad \dot{b}_i=R_i\nu_i.\end{equation} The potential energy for the $i^{th}$ agent is denoted by $U_i(b_i,R_i):SE(3)\to\mathbb{R}$. Then the Lagrangian for the motion of the $i^{th}$ rigid body, after a left-trivialization, $\mathcal{\ell}_i:SE(3)\times\mathbb{R}^{6}\to\mathbb{R}$ is given by $$\ell_i(b_i,R_i,\nu_i,\Omega_i)=\frac{1}{2}\langle J_i\Omega_i,\Omega_i\rangle+\frac{1}{2}\langle M_i\nu_i,\nu_i\rangle-U_i(b_i,R_i),$$ where $\langle\cdot,\cdot\rangle$ is the trace pairing (an inner product) given by $\langle A,B\rangle:=\hbox{Tr}(A^{T}B)$, $J_i$ the inertia matrix and $M_i$ the mass matrix for the $i^{th}$ rigid body.

We assume that each vehicle is fully actuated, where control acts on the dynamics. The controlled dynamics of each vehichle is determined by Euler-Lagrange equations \eqref{eq_ep_intro} with controls, for the Lagrangian $\ell_i$, i.e., equations \eqref{eq:left-invariant-dynamical-system-multi-agents}. In this context, equations \eqref{eq:left-invariant-dynamical-system-multi-agents} are given by
\begin{align}
M_i\dot{\nu}_i=&M_i\nu_i\times\Omega_i+\mathcal{U}_i(b_i,R_i)+u_i,\label{equ1}\\
J_i\dot{\Omega}_i=&J_i\Omega_i\times\Omega_i+M_i\nu_i\times\nu_i+\mathcal{W}_i(b_i,R_i)+\overline{u}_i,\label{equ2}
\end{align}together with \eqref{kinematicSE(3)}, where $u=(u_1,u_2,u_3)$, $\bar{u}=(u_4,u_5,u_6)\in\mathbb{R}^{3}$ and $\mathcal{U}_i(b_i,R_i)$, $\mathcal{W}_i(b_i,R_i)\in\mathbb{R}^{3}$ are defined by   \begin{align}\mathcal{U}_i(b_i,R_i):=&-R_i^{T}\frac{\partial U_i}{\partial b_i}(b_i,R_i),\label{control1}\\ \widehat{\mathcal{W}}_i(b_i,R_i):=&\frac{\partial U_i^{T}}{\partial R_i}R_i-R_i^{T}\frac{\partial U_i}{\partial R_i}\label{control2}.\end{align}
We assume that each agent occupies a sphere $S_i=\{b\in\mathbb{R}^{3}:||b-b_i||\leq r_i\}$ where $b_i\in\mathbb{R}^{3}$ coincides with the center of the sphere and $r_i$ its radius.

The feasibility for the coordinated motion is completely specified by the (holonomic) collision avoidance  constraints $\phi_{12}^{1}$, $\phi_{13}^{2}$  and $\phi_{23}^{3}$, by giving a prescribed distance $d_{ij}^{k}\in\mathbb{R}^{+}$ between the center of masses of the bodies.
 
The set of constraints $\mathcal{C}=\{\phi_{12}^{1}, \phi_{13}^{2}, \phi_{23}^{3}\}$ is determined by \begin{equation}\label{constrainSE(3)}\phi_{ij}^{k}(g_i,g_j)=||b_i-b_j||^{2}-(r_i+r_j+d_{ij}^{k})^2=0\end{equation} and specifies the functions $\Phi_{ij}^{k}:SE(3)^{2}\to\mathbb{R}$. Consider the vector valued function $\Phi:SE(3)^{3}\to\mathbb{R}^{3\times 1}$ given by $\Phi(g)=[\Phi_{12}^{1}(g), \Phi_{13}^{2}(g), \Phi_{23}^{3}(g)]^{T}$, where $g=(g_1,g_2,g_3)$.Denoting by \hbox{``grad' }the gradient of functions, a simple computation show that $$T_{g_i}L_{g_i^{-1}}(\hbox{grad }\Phi_{ij}^{k})=2(0,R_i^{T}(b_i-b_j),0,-R_i^{T}(b_i-b_j)).$$ Consider the augmented Lagrangian $\overline{L}:SE(3)^3\times \mathfrak{se}(3)^3\times\mathbb{R}^{3}\to\mathbb{R}$ with $\lambda=(\lambda_1,\lambda_2,\lambda_3)\in\mathbb{R}^{3}$ given by \begin{align*}\overline{L}(b,R,\nu,\Omega,\lambda)=&\sum_{i=1}^{3}\ell_i(b_i,R_i,\nu_i,\Omega_i)-\frac{1}{2}\lambda_1\phi_{12}^{1}(g_1,g_2)\\&-\frac{1}{2}(\lambda_2\phi_{13}^{2}(g_1,g_3)+\lambda_3\phi_{23}^{3}(g_2,g_3)).\end{align*}
By Theorem \ref{theorem1var}, the set of differential equations for the feasibility in the coordinated motion are given by \begin{align}
M_1\dot{\nu}_1=&M_1\nu_1\times\Omega_1+\mathcal{U}_1(b_1,R_1)\label{a1}\\&+\lambda_1R_1^{T}(b_1-b_2)+\lambda_2R_1^{T}(b_1-b_3)\nonumber\\
M_2\dot{\nu}_2=&M_2\nu_2\times\Omega_2+\mathcal{U}_2(b_2,R_2)\label{a2}\\&+\lambda_1R_2^{T}(b_1-b_3)+\lambda_3R_2^{T}(b_2-b_3)\nonumber\\
M_3\dot{\nu}_3=&M_3\nu_3\times\Omega_3+\mathcal{U}_3(b_3,R_3)\label{a3}\\&+\lambda_2R_3^{T}(b_1-b_3)+\lambda_3R_3^{T}(b_2-b_3)\nonumber\\
J_i\dot{\Omega}_i=&J_i\Omega_i\times\Omega_i+M_i\nu_i\times\nu_i+\mathcal{W}_i(b_i,R_i),\label{a4}
\end{align} for $i=1,2,3$, together with equations \eqref{constrainSE(3)} and \eqref{kinematicSE(3)}.

In some rigid body applications as for instance spacecraft motion on $SO(3)$, the mass matrix is usually given by $M_i=m_iI_i$ where $m_i$ is the mass of the body and $I_i$ its matrix of inertia moments. We will consider models for underwater vehicles where the elements of $M_i$ may be different due to the fact that added masses have to be taken into account.

For simplicity in this expository modelling for our theoretical results, we assume that possible dissipative forces acting on the body under the water are 
negligible. The potential energy for the $i^{th}$ underwater vehicle is given by $$U_i(R_i,b_i)=\rho\gamma_i g\langle \bar{r}_i,R_i^{T}e_3\rangle+(\rho\gamma_i-m_i)gb_i^{z},$$ where $g$ is the gravitational acceleration, $m_i$ are the masses of each body, $\rho$, 
is the density of water, $\gamma_i$ is the volume of each body, and $\bar{r}_i\in\mathbb{R}^3$ is a vector from the center of gravity to the center of buoyancy (in the body fixed frame) of each body. The positive $z$-axis in $\mathbb{R}^3$ for each body, i.e., $b_i^{z}$, is
taken to point downwards in the same direction as the gravity. Under these considerations, equations \eqref{a1}-\eqref{a4} are given by \begin{align}
M_1\dot{\nu}_1=&M_1\nu_1\times\Omega_1-R_1^{T}(m_1-\rho\gamma_1)ge_3\nonumber\\&+\lambda_1R_1^{T}(b_1-b_2)+\lambda_2R_1^{T}(b_1-b_3)\label{eqqf1}\\
M_2\dot{\nu}_2=&M_2\nu_2\times\Omega_2-R_2^{T}(m_2-\rho\gamma_2)ge_3\nonumber\\&+\lambda_1R_2^{T}(b_1-b_3)+\lambda_3R_2^{T}(b_2-b_3)\label{eqqf2}\\
M_3\dot{\nu}_3=&M_3\nu_3\times\Omega_3-R_1^{T}(m_3-\rho_3\gamma)ge_3\nonumber\\&+\lambda_2R_3^{T}(b_1-b_3)+\lambda_3R_3^{T}(b_2-b_3)\label{eqqf3}\\
J_i\dot{\Omega}_i=&J_i\Omega_i\times\Omega_i+M_i\nu_i\times\nu_i-\rho\gamma_i g\bar{r}_i\times(R^{T}_ie_3),\label{eqqf4}
\end{align} for $i=1,2,3$, together with equations \eqref{constrainSE(3)} and \eqref{kinematicSE(3)}.
\subsection{Construction of control law for the cooperative motion}
The step-by-step algorithm to construct the control law is summarized in Algorithm \ref{Algorithm}. \begin{algorithm}[h!]
\small
\caption{Construction of the control law for the cooperative motion}
\label{Algorithm}
\begin{algorithmic}[1]
\State \textbf{Data:} $M_i,J_i,\bar{r}_i, r_i, \rho,\gamma_i,g, d_{ij}^{k}$, time step $h$, $\#$ of steps $N$.\\ \textbf{inputs:} $R_i(0)$, $\Omega_i(0)$, $b_i(0)$, $\nu_i(0)$, $i=1,2,3$, satisfying  the constraints \eqref{constrainSE(3)} and regularity condition \eqref{regularity}, $T=Nh$.

\Comment \emph{first stage: Dynamics of $\lambda(t)$}
          
           \State Compute the derivative w.r.t. time in equation \eqref{constrainSE(3)}. For the derivative w.r.t time of each constraint \eqref{constrainSE(3)}, isolate $\dot{b}_i$ and replace into the obtained expression $\dot{b}_j$  from \eqref{kinematicSE(3)}.
         
                \State Compute the second derivative w.r.t. time in equation \eqref{constrainSE(3)}. For each $i$ isolate $\ddot{b}_i$, and in the isolated expression replace $\dot{b}_i$ from Step 3 and $\dot{\nu}_i$ from \eqref{eqqf1}-\eqref{eqqf3}.\State From the expression obtained in Step 4, isolate $\lambda(t)$ as a function of $\dot{b}$ and $\ddot{b}$.
\State Use Steps $3$ and $4$ of the algorithm to write derivatives of $b$ in the expression of $\lambda(t)$ obtained by Step 5, in terms of configurations and obtain the expression for the evolution of $\lambda(t)$ in terms of $b_i$, $\nu_i$ and $R_i$.
\State Replace $\lambda(t)$ in terms of $b_i$, $\nu_i$ and $R_i$ obtained in Step $6$ in equations \eqref{eqqf1}-\eqref{eqqf3}. \\
   \Comment \emph{second stage: Solve the equations \eqref{eqqf1}-\eqref{eqqf4}}
    \For {$i =1 \to 3$} 
                \State solve \eqref{eqqf1}-\eqref{eqqf4} subject to \eqref{kinematicSE(3)}.
    \EndFor
    \State \textbf{outputs:} $R_i(t)$, $b_i(t)$, $\Omega_i(t)$, $\nu_i(t)$ for $i=1,2,3$.

 \Comment \emph{third stage: Construction of the control law}

    \For {$i =1 \to 3$}   
            \State Replace $R_i(t)$, $b_i(t)$, $\Omega_i(t)$, $\nu_i(t)$,into \eqref{equ1}-\eqref{control2} and solve for $u_i(t)$ and $\bar{u}_i(t)$.
  
    \EndFor
\State \textbf{outputs:} $u_i(t)$ and $\bar{u}_i(t)$ from equations \eqref{equ1} and \eqref{equ2}.
\Statex
\end{algorithmic}
 \end{algorithm}
\subsection{Simulation results}

\begin{figure}[h!]
\includegraphics[width=.4\linewidth]{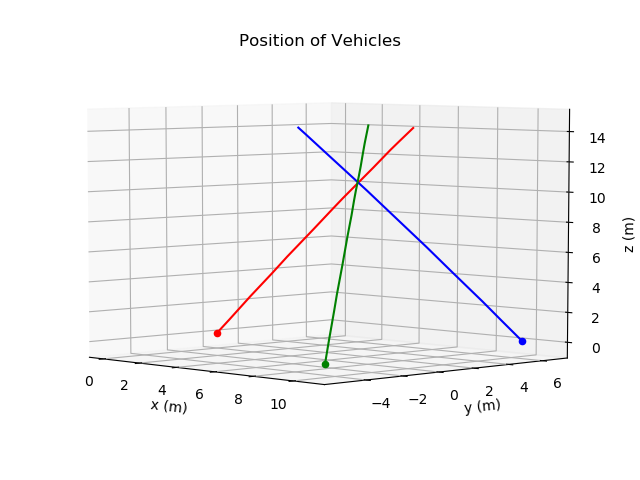}
  \centering
  \includegraphics[width=.4\linewidth]{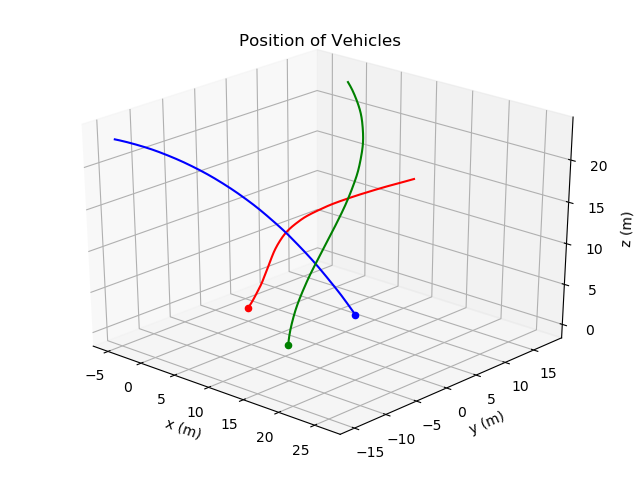}
\caption{Collision of vehicles vs. collision avoidance.}
\label{fig5}
\end{figure}

\begin{figure}[h!]
\includegraphics[width=.32\linewidth]{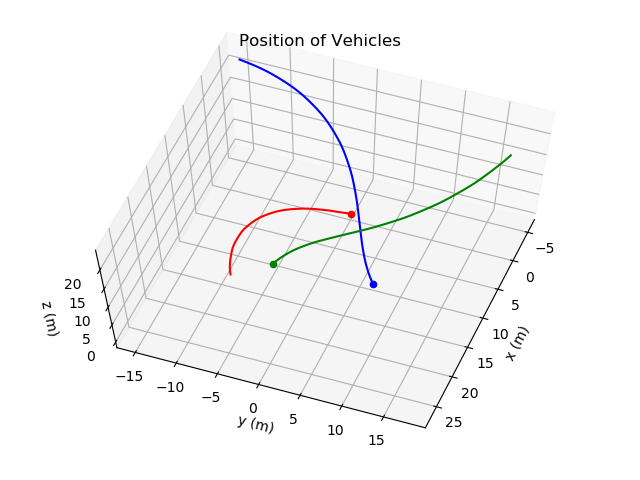}
  \centering
  \includegraphics[width=.32\linewidth]{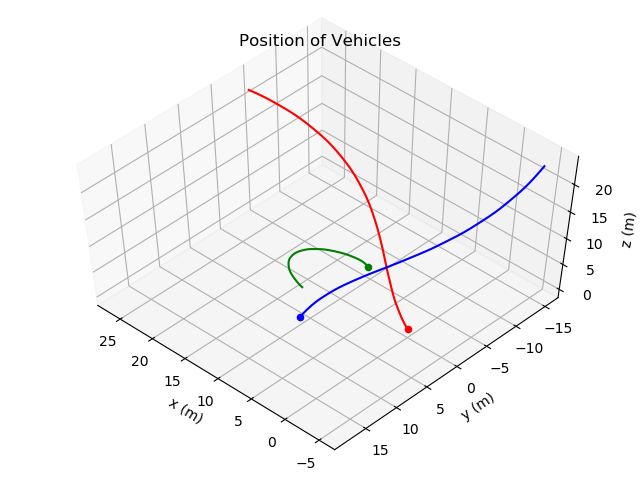}
 \centering
  \includegraphics[width=.32\linewidth]{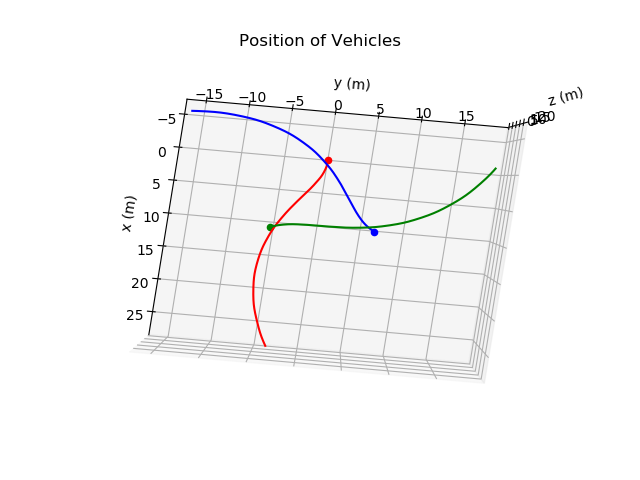}
\caption{Different perspectives of the trajectories for collision avoidance.}
\label{fig6}
\end{figure}

\begin{figure}[h!]
\includegraphics[width=.31\linewidth]{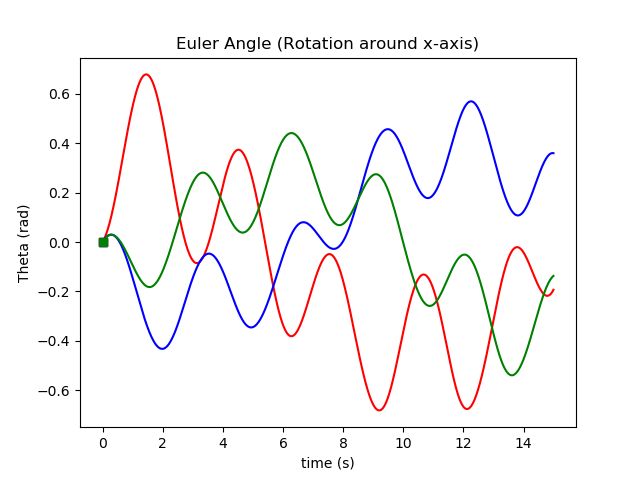}
  \centering
  \includegraphics[width=.31\linewidth]{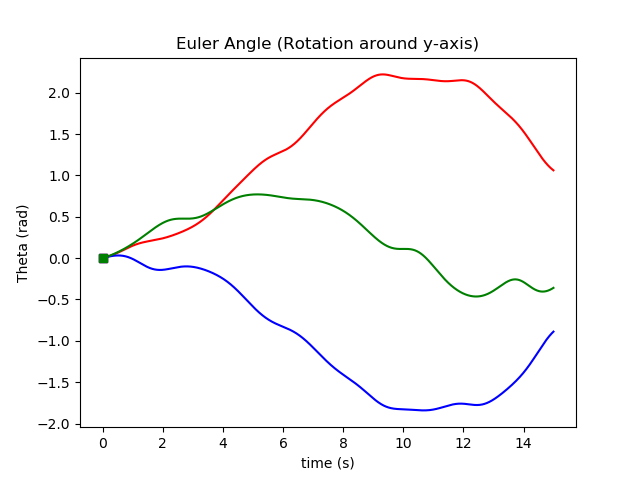}
 \centering
  \includegraphics[width=.31\linewidth]{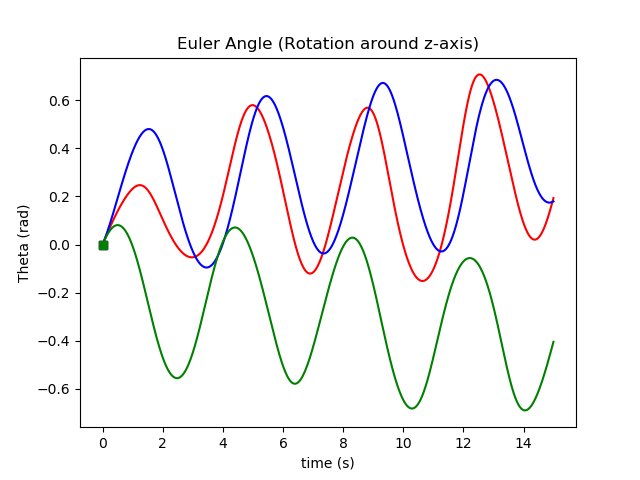}
\caption{Evolution of Euler angles in the rotation matrix $R(t)$.}
\label{fig7}
\end{figure}

\begin{figure}[h!]
\includegraphics[width=.31\linewidth]{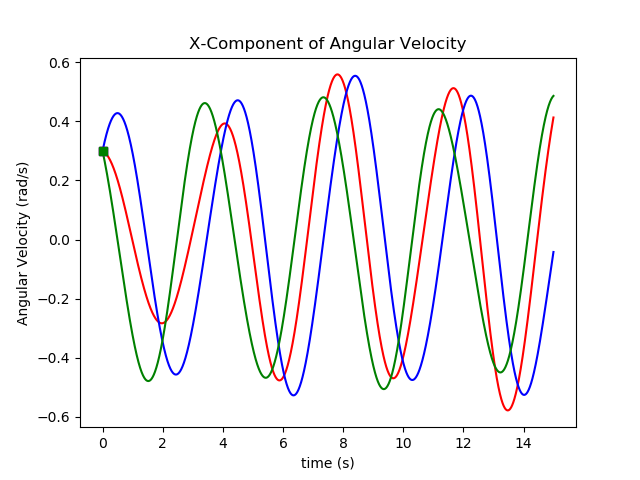}
  \centering
  \includegraphics[width=.31\linewidth]{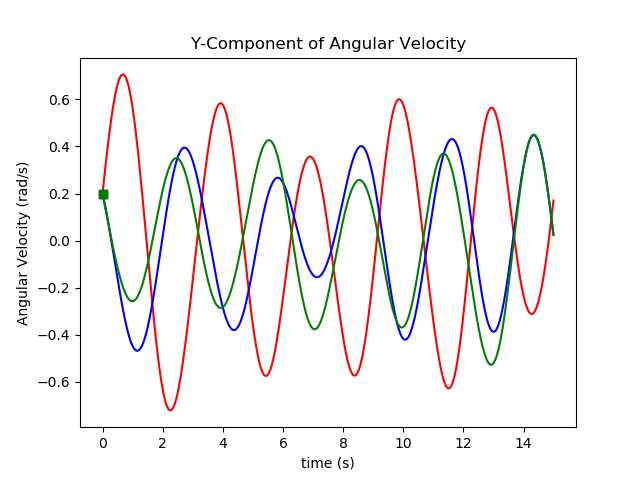}
 \centering
  \includegraphics[width=.31\linewidth]{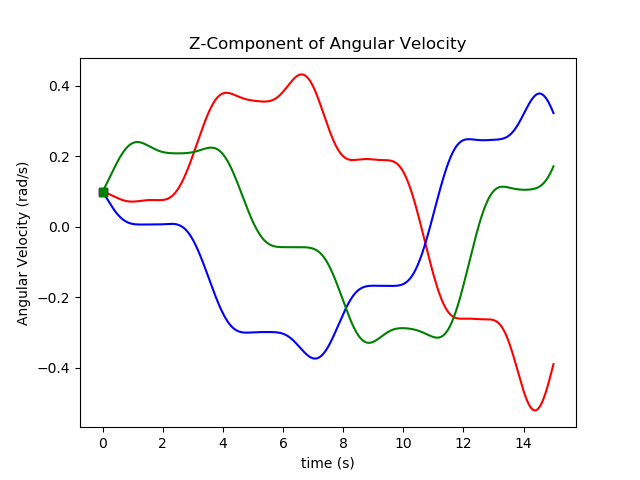}
\caption{Evolution of angular velocity $\Omega(t)$.}
\label{fig8}
\end{figure}

Now we show how the previous algorithm is employed in numerical simulation. We consider that the three bodies have mass $m_i = 123.8$kg,
and mass (including added masses) and inertia matrices 
$M_i=m_iI_i + \hbox{diag}(65, 70, 75)$kg,
$J_i=\hbox{diag}(5.46, 5.29, 5.72)$kg$\times$ m$^2$ and $I_i=\hbox{Id}_{3\times 3}$kg$\times$ m$^2$, with $\hbox{Id}_{3\times 3}$ the $(3\times 3)$-identity matrix. Also assume that $\rho\gamma_i g=1215.8$N and 
$\bar{r}_i = (0, 0,-0.007)^{T}$m. Initial conditions are chosen as $R_i(0)=\hbox{Id}_{3\times 3}$s$^{-1}$, $\Omega_i(0)=(0.3,0.2,0.1)^{T}$s$^{-1}$, $\nu_i(0)=R_i(0)^{-1}(0.1,0.2, 1)^{T}$ms$^{-1}$, $d_{12}^{1}=d_{13}^{2}=d_{23}^{3}=10$m, $b_1= (0, 0, 0)^{T}$m, $b_2 = (10, 6.63324958, 0)^{T}$m, $b_3 = (10.7446, -5.34363, 0)^{T}$m. The radius of the spheres $r_i$ which contains each body is $1$m.  With the above choice of parameters and initial conditions satisfying the constraints and the regularity condition \eqref{regularity} we simulate the controlled dynamics of the  vehicles with a step size of $h = 0.005$s using an Euler method. In Figure \ref{fig5} we compare the position of the center of mass of the bodies without the collision avoidance constraints (left) and with our method (right) for $N=5000$. We observe that trajectories crosses each others without our method, while the avoidance of trajectories crossing each others occurs when we incorporate the collision avoidance constraints \eqref{constrainSE(3)}. In Figure \ref{fig6} we show different perspectives for the collision avoidance trajectory. Figures \ref{fig7} and \ref{fig8} show the attitude and angular velocity, respectively, of the three bodies with the collision avoidance constraints.

\begin{IEEEbiography}[{\includegraphics[width=1in,height=1.25in,clip,keepaspectratio]{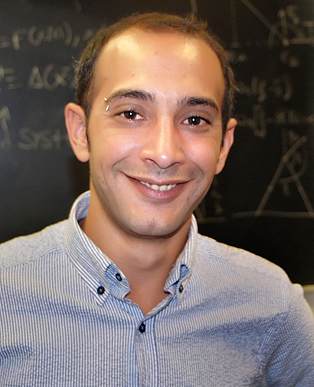}}]{Leonardo J. Colombo.}
Leonardo Colombo is a Postdoctoral Junior Leader from La Caixa Foundation at Instituto de Ciencias Matematicas  (ICMAT) from 2019. He received a B.Sc. in Mathematics from the Universidad Nacional de La Plata, Argentina in 2009, a M.S in Applied Mathematics from the Universidad Aut\'onoma of Madrid (UAM) in 2012, and a Ph.D. from ICMAT and UAM in 2014. His research interest includes multi-agent control systems, geometric mechanics, geometric integration, hybrid systems, robot motion planning, unmanned aerial vehicles, formation control and control oriented learning.

He was a Postdoc Assistant Professor at University of Michigan, USA from 2014 to 2017, a Postdoctoral researcher at ACCESS LINEAUS Center, Department of Automatic Control, School of Electrical Engineering, KTH Royal Institute of Technology, Stockholm,
Sweden between 2017 and 2018 and a Juan de la Cierva Incorporaci\'on researcher at ICMAT from 2018 to 2019. He received the Vicent Casseles Award from the Spanish Royal Mathematical Society and Foundation BBVA in 2016.
\end{IEEEbiography} \begin{IEEEbiography}[{\includegraphics[width=1in,height=1.25in,clip,keepaspectratio]{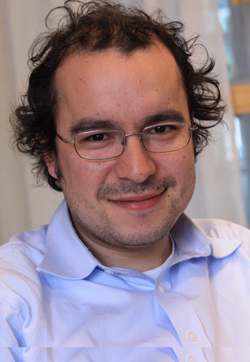}}]{Dimos V. Dimarogonas.}
Dimos V. Dimarogonas received the Diploma in Electrical
and Computer Engineering in 2001 and the Ph.D. in
Mechanical Engineering in 2007, both from the National
Technical University of Athens (NTUA), Greece. From May
2007 to February 2009, he was a Postdoctoral Researcher
at the Automatic Control Laboratory, School of Electrical
Engineering, KTH Royal Institute of Technology, Stockholm,
Sweden, and a Postdoctoral Associate at the Laboratory
for Information and Decision Systems, Massachusetts
Institute of Technology (MIT), Cambridge, MA, USA. He
is currently an Associate Professor in Automatic Control,
School of Electrical Engineering, KTH Royal Institute of Technology. His current
research interests include multi-agent systems, hybrid systems, robot navigation,
networked control and event-triggered control.
Dr. Dimarogonas was awarded a Docent in Automatic Control from KTH in
2012. He serves on the Editorial Board of Automatica, the IEEE Transactions on Automation Science and Engineering and the IET Control Theory and Applications,
and is a member of the Technical Chamber of Greece. He received an ERC Starting
Grant from the European Commission for the proposal BUCOPHSYS in 2014 and was
awarded a Wallenberg Academy Fellow grant in 2015.

\end{IEEEbiography}
\end{document}